\pdfoutput=1
\documentclass[sigconf,screen]{acmart} 
\usepackage{amsmath,amssymb,amsfonts}
\usepackage{graphicx}
\usepackage{textcomp}
\usepackage{array}
\usepackage{booktabs} % For formal tables
\usepackage[breakable]{tcolorbox}
\tcbuselibrary{hooks, breakable, skins}
\usepackage{tabularx}
\usepackage{multirow}
\usepackage[flushleft]{threeparttable}
\usepackage{amssymb}
\usepackage{enumitem}
\usepackage{rotate}
\usepackage{graphicx}
\usepackage{pdfpages}
\usepackage{colortbl}
\usepackage{arydshln}
\usepackage{pifont}
\usepackage{times}
\usepackage{rotating}
\usepackage{makecell} 
\usepackage{tabularx}
\usepackage{balance}
\usepackage{booktabs}
\usepackage{wrapfig}
\usepackage[pdf]{graphviz}
\usepackage{tikz}
\usetikzlibrary{angles, positioning}
\usepackage{makecell}
\usepackage{xcolor}
\usepackage{soul}
\usepackage{tabu}
\usepackage{dblfloatfix} 

\usepackage{tikz}
\def\checkmark{\tikz\fill[scale=0.4](0,.35) -- (.25,0) -- (1,.7) -- (.25,.15) -- cycle;}

\usepackage{graphicx} %Loading the package
\graphicspath{{imgs/}}

\newcommand{\bi}{\begin{itemize}}
\newcommand{\ei}{\end{itemize}}
\newcommand{\be}{\begin{enumerate}}
\newcommand{\ee}{\end{enumerate}}
\newcommand{\fig}[1]{Figure~\ref{fig:#1}}
\newcommand{\tab}[1]{Table~\ref{tab:#1}}

\usepackage{amsmath}
\usepackage{subfigure}
\usepackage{url}
\usepackage{mathtools}
\usepackage{amsmath}
\usepackage[final]{listings}
\renewcommand{\textrightarrow}{$\rightarrow$}
\usepackage{algorithm}
\usepackage[noend]{algpseudocode}

\begin{document}

\title{Predicting Breakdowns in Cloud Services (with SPIKE)}
%  \author[J. Chen]{Jianfeng Chen, Joymallya Chakraborty\\Tim Menzies}
% \affiliation{
% \department{Computer Science, NC State University, USA}}
% \email{{jchen37,jchakra}@ncsu.edu;timm@ieee.org}

% \author[P. Clark et al.]{Philip Clark, Kevin Haverlock, Snehit  Cherian  }
% \affiliation{%
% \department{LexisNexis Legal \& Professional, USA}}
% \email{{philip.clark,kevin.haverlock,snehit.cherian} }  % DO NOT REMOVE SPACE BETWEEN } }
% \email{@lexisnexis.com}
\author{Jianfeng Chen}
\email{jchen37@ncsu.edu}
\orcid{0000-0002-4668-3412}
\affiliation{%
  \institution{NC State University}
  \streetaddress{890 Oval Drive}
  \city{Raleigh}
  \state{NC}
  \postcode{27695}
  \country{USA}
}

\author{Joymallya Chakraborty}
\email{jchakra@ncsu.edu}
\affiliation{%
  \institution{NC State University}
  \streetaddress{890 Oval Drive}
  \city{Raleigh}
  \state{NC}
  \postcode{27695}
  \country{USA}
}

\author{Philip Clark}
\email{philip.clark@lexisnexis.com}
\affiliation{%
  \institution{LexisNexis Legal \& Professional}
  \city{Raleigh}
  \state{NC}
  \country{USA}
}

\author{Kevin Haverlock}
\email{kevin.haverlock@lexisnexis.com}
\affiliation{%
  \institution{LexisNexis Legal \& Professional}
  \city{Raleigh}
  \state{NC}
  \country{USA}
}

\author{Snehit Cherian}
\email{snehit.cherian@lexisnexis.com}
\affiliation{%
  \institution{LexisNexis Legal \& Professional}
  \city{Raleigh}
  \state{NC}
  \country{USA}
}

\author{Tim Menzies}
\email{timm@ieee.org}
\orcid{0000-0002-5040-3196}
\affiliation{%
  \institution{NC State University}
  \streetaddress{890 Oval Drive}
  \city{Raleigh}
  \state{NC}
  \postcode{27695}
  \country{USA}
}

\renewcommand{\shortauthors}{Chen and Clark, et al.}

\begin{abstract}

 Maintaining   web-services  is a 
mission-critical task where
any downtime   means loss of revenue and
  reputation (of being
  a reliable service provider).
In the current
 competitive web services market, such a loss of reputation
 causes extensive loss of future revenue.

To address this issue, we developed {\it SPIKE},
a data mining tool which can predict 
upcoming service breakdowns,  half an hour
into the future. 
Such predictions let an organization alert and assemble the tiger team to address the problem (e.g. by 
  reconfiguring   cloud hardware in order to reduce
  the likelihood of that breakdown).
 
{\it SPIKE} utilizes
(a)~regression tree learning (with CART);
(b)~synthetic minority over-sampling (to handle how rare spikes are in our 
data); (c)~hyperparameter optimization (to learn best settings for our
local data) and (d)~a technique we called ``topology sampling''
where training vectors are built from extensive details of an individual node
plus summary details on all their neighbors.

In the experiments reported here,   
{\it SPIKE} predicted service
spikes  30  minutes into future
with   recalls and precision of   75\% and above.
Also,
{\it SPIKE}
 performed relatively  better
than   other widely-used learning methods  (neural nets,
random forests, logistic regression).

\end{abstract}
 



%%% The following is specific to ESEC/FSE '19-IND and the paper
%%% 'Predicting Breakdowns in Cloud Services (with SPIKE)'
%%% by Jianfeng Chen, Joymallya Chakraborty, Philip Clark, Kevin Haverlock, Snehit Cherian, and Tim Menzies.
%%%
\setcopyright{acmcopyright}
\acmPrice{15.00}
\acmDOI{10.1145/3338906.3340450}
\acmYear{2019}
\copyrightyear{2019}
\acmISBN{978-1-4503-5572-8/19/08}
\acmConference[ESEC/FSE '19]{Proceedings of the 27th ACM Joint European Software Engineering Conference and Symposium on the Foundations of Software Engineering}{August 26--30, 2019}{Tallinn, Estonia}
\acmBooktitle{Proceedings of the 27th ACM Joint European Software Engineering Conference and Symposium on the Foundations of Software Engineering (ESEC/FSE '19), August 26--30, 2019, Tallinn, Estonia}

\settopmatter{printfolios=true}
\maketitle

\section{Introduction}

Managing cloud services is an important problem. Mismanaging such services
 results in server downtime and an associated loss of revenue, particularly
for organizations with  penalty clauses in their service contracts.
Even down times of just a few hours each month can be  detrimental to the professional reputation of an cloud-service provider. This is a concern since organizations with a poor
reputation for reliability have a harder time attracting and retaining clients.

\begin{figure}[!b]
  \includegraphics[width=.9\linewidth]{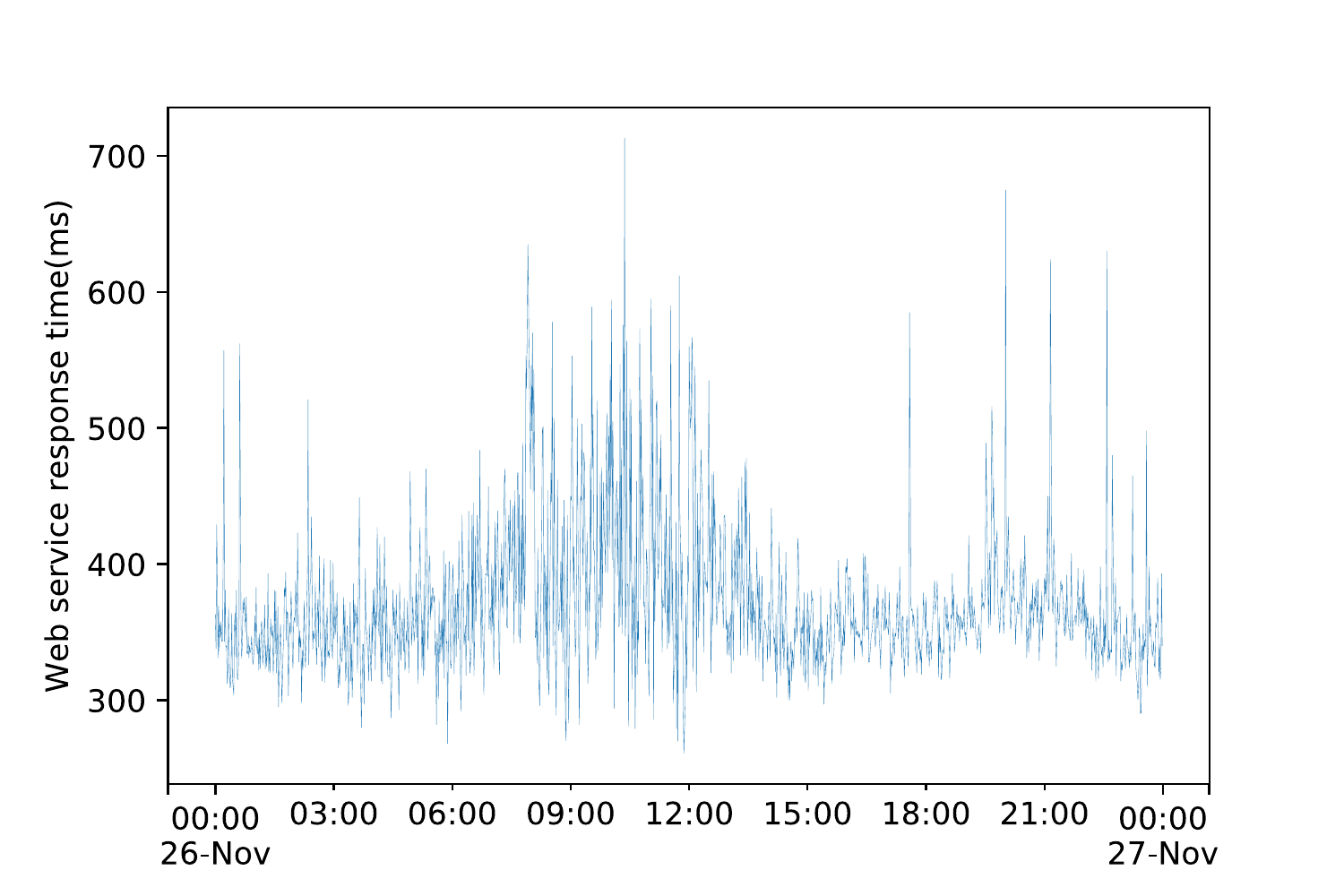}
  \caption{Web service response time; one cloud compute node, 11/26/2018.}
  \label{fig:spike}
\end{figure}

This
paper explores one  kind of breakdown-- specifically, service spikes
that can freeze up a cloud server. In  \fig{spike}, such spikes are clearly visible (see the higher values).
At first glance, such spikes seem relatively infrequent and very small (the y-axis of \fig{spike} is in milliseconds). However, it should be remembered that a modern web page shows results from dozens of microservices, each of which uses dozens of other queries 
to the underlying databases.  Spikes like those shown in
 \fig{spike}
can lead to frustratingly slow systems performance (e.g. very slow displays of new web pages).  Hence, such spikes are critical  business events  that can damage an organization's reputation as a reliable cloud service provider.

Predicting service spikes is hard since they can occur rarely and may occur as sudden
extreme outliers. For example, in \fig{spike}, the large 11am and 8:30pm spikes might be anticipated by the steady
build-up in the proceeding hour. However, between 6pm and 3am, we count six other spikes that are not proceeded by
any apparent build-up.

Another factor complicating spike prediction 
is  the  rapidly
changing nature of     cloud environments.
For example, consider LexisNexis
(the organization that funded this research).
In the   year or two, LexisNexis has  retired its locally managed CPU farms in favor
of CPU farms managed by  multiple major cloud vendors.
% (including Azure\footnote{The Microsoft cloud service.} and AWS\footnote{Amazon Web Services}).
But cloud instance managements tools 
% (such as   Kubernetes
% or Docker SWARM) 
are   evolving rapidly. Hence, like every other user of
cloud-based services, LexisNexis anticipates that, in the near
future, it will change its web architecture yet again.
% (e.g. to  Kubernetes clusters). 

Due to this relentless pace of change, much of 
the prior operational history is not relevant to
current or future operations. Hence, methods that work in prior 
studies may not work for future studies.  This means that data science teams working on cloud clusters
are forced to constantly update their model.

Further, at each update, new learning technologies may
be necessary. For example, the authors of this paper started with established methods for predicting service spikes~\cite{syu2018survey} and  when those did not work, we moved on to   other methods 
(described later in this paper). In all, we spent three months building and discarding a dozen different predictors\footnote{During this trial-and-error period,
we often took solace from Thomas Edison's famous quote ``I have not failed. I've just found 10,000 ways that won't work.''} before finding one that could handle the specifics of the LexisNexis environment.

Initially, we imagined that we would be building a recommender system
that would suggest the number and type of cloud server instances that should be added or deleted 
in order to maintain service availability (at minimum cost).  In theory, such a recommender system could
be learned from the historical logs of prior nominal and off-nominal behavior.

However, once
 we realized how
fast the cloud services were changing, we also realized that much of the historical log was no
longer relevant to current practice.
So  we changed track and asked ``what are the major pain points of running the LexisNexis
cloud service?''. 
This new question prompted our  subject matter experts to
recounted various war stories about what happens
when a service spike occurs. One issue with those events is gathering together the response team. ``It can take
five to ten minutes to realize we have a problem'', we were told, ``after which it can take another
few minutes of calling/texting to get everyone we need into  a conference call''.

% \begin{figure*}[h]
% \includegraphics[width=10cm]{architecture.jpg}
% \caption{Architecture of LexisNexis Search \& Retrieval System. Each of the boxes here has its own complex
% internal structure. Our task is to understand the net result of all these components working together, on the cloud.}
% \label{fig:architecture}
% \end{figure*}

From remarks such as this, we realized our goal needed to be  ``early warning''. Accordingly,
the goals of this project were set to be:
\begin{quote}
{\em Build \underline{comprehensible} and 
\underline{effective} predictors for service spikes, 30 minutes into the future.}
\end{quote}
As to what constitutes a ``service spike'', using the results of Section~4, we defined
that to be values over  470 ms/query. 
Such a predictor would allow an organization
to reduce their response time to forthcoming incidents. Further, in some
cases, it would be possible to {\em remove} the cause the spike, thus 
preventing that incident from  occurring in the first place.

Note that the above goal includes  {\em comprehensible models}. 
Our experts required some report of the lessons learners that they can read, understand and audit.
Hence, we need to use data mining methods
that produce human readable models (e.g. not Naive Bayes classifiers, not neural networks, not instance-based learners, not random forests).

\subsection{Organization of this Paper}

The rest of this paper explores methods for building comprehensible and effective predictors for service spikes
of over 470 milliseconds per query, 30 minutes into the future. Our paper is structured as follows. 
Section 2 first briefly introduces some background of the LexisNexis information retrieval system, features of search engine we used, as well as the motivations of the project.
Section 3 goes over all machine learning techniques  explored during the project.
Section 4 presents the data we collected for the prediction in the project.
Section 5 reveals the exploration of all steps, leading to the managerial or technical summary.
Finally we conclude the project in section 6 as well as the future work in section 7.

\section{Background}
\subsection{Business Context}\label{sect:motivation}
LexisNexis  provides computer-assisted legal research, business research and risk management services \cite{lnnyt,lnnyt1}. During the 1970s, LexisNexis pioneered  electronic accessibility of legal and journalistic documents~\cite{lnnyt2}. Since  2006, it has the world's largest electronic database for legal and public-records  information \cite{lnnyt2}.

LexisNexis provides regulatory, legal, business information and analytics to the legal community.  Legal and research professionals use the LexisNexis Advance platform to find relevant information~\cite{lnlegal} that they use to prepare legal cases and drive better legal outcomes. LexisNexis operates across multiple major cloud vendors.
% (including Azure and AWS)
As a large distributed system, LexisNexis is mindful of costs and wants to scale the database and search service when demand is anticipated or predicted to be needed.  Hence, this paper.

\subsection{System Architecture}

The LexisNexis  database contains over  100 billion documents and records.  Records are added at the rate of
nearly two million new items daily
from over 50,000 sources.  In all, over 20 million legal documents are processed daily from legal jurisdictions throughout the world.  In addition, the databases contain over 320 million company profiles with content archives dating back 40 years. 

To support all this, 
the Lexis Advanced product suite is a modern web application consisting of a central monolithic application, supported by hundreds of micro-services.
% {\color{red} Figure \ref{fig:architecture} illustrates the architecture of the search \& retrieval system.}
Individual micro-services perform a wide array of functions, including content loading and enrichment, document search and retrieval, user authentication, and notifications.  Many of these services are shared by multiple other services, forming an interconnected web of dependencies.  When service disruptions occur, they are felt as a cascade of failures in which one service effectively disables multiple other services.

LexisNexis utilizes a number of technologies to constantly monitor the state of the application environment.  Application logs, server logs, host metrics, web traffic, infrastructure status, and user activity are all monitored using various automated tools
% such as Splunk\footnote{ Splunk is an  engine for monitoring, searching, analyzing, visualizing and acting on voluminous streams of real-time machine data.}, New Relic\footnote{New Relic monitors web and mobile applications in real-time  with support for custom-built plugins to collect arbitrary data.}, CloudWatch\footnote{  CloudWatch is  AWS'  monitoring and management service built for developers, system operators, site reliability engineers (SRE), and IT managers.}, 
and in-house software. 
One of them \footnote{In this paper, some  terms are anonymized for proprietary business reasons.} is primarily used to ingest, parse, and visualize data from log files, while the other tool is used to monitor application performance metrics such as response time and throughput.  Both tools allow users and automated scripts to monitor and react to quickly changing conditions in near real time. 
% Figure \ref{fig:architecture} shows the overall architectural view of the search and retrieval system.

%\subsection{Data Collection}\label{sect:dc}

% From these soruces,  data was collected from New Relic or Splunk, and consisted of application response time, throughput, error rate, and Apdex (the  metric calculated from response time that is intended to highlight user experience). We jointly spent three months to collect the data which indicates the complexity and effort of collecting such metric data.

\subsection{Document Storage and Search}
Currently, LexisNexis makes extensive use of  
% MarkLogic
some
multi-model NoSQL  database 
for document storage and searching.
% ~\cite{marklogicReview}. 
% With over two decades of development experience, MarkLogic servers are used in  thousands of commercial mission-critical environments around the world. 
Such document search engine stores and queries data in documents, graph data, or relational data, providing incredible flexibility.
It  includes  a search engine that is especially suitable for full-text searches (most documents in LexisNexis are in full-text).

The database in LexisNexis  supports  massive horizontal and can support  installations
with hundreds of nodes, petabytes of data, and billions of documents
(which still processing tens of thousands of transactions per second).
The system also is a \textit{trusted database} having all the enterprise features required to handle sensitive enterprise data:

\bi
\item \textbf{Advanced Security}: It  offers granular security controls at the document and even element/property level, redaction and anonymization for safe sharing, as well as advanced encryption. 
% MarkLogic is one of the few cloud-scale
% databases with  a common criteria security certification.
\item \textbf{ACID Transactions}: It has multi-document ACID transactions that provide data consistency even with large-scale transaction applications.
\item \textbf{Cloud Neutral}: It has been successful running in the cloud for over a decade and it is compatible with any public cloud provider.
\ei

However, such trusted database can lead to operational issues.
At its core, it is a {\em twin instance} system where data is stored on Node1 and indexed on Node2. This twin architecture
increases survivability of the system against insult (if the index node goes down, it can be rebuilt elsewhere). 
On the other hand, experience has shown that this twin model
can complicate the operation of such database system.
Nodes cannot be simply added (if more performance is required) or removed (if we want to save operational costs if the CPUs are under-utilized). There is also an additional operational cost of such a system-- after a crash (when data has to be rebuilt), some expert  human supervision is required
to appropriately partition the data across the servers.

In the meanwhile, LexisNexis may soon be running its database servers on some open-sourced container  orchestration -- at which time,
the expertise needed to run a database server will need to be updated, yet again. 
Because of this relentless pace of change in cloud services. we made the design decision to build {\it SPIKE} without using
detailed knowledge of the current internal database system. Instead, as discussed in section \ref{sect:data},
when we  did data collection, we restricted ourselves to measurements we might reasonably expect to see in a wide range of future cloud environments.

% \tosign[kevin]{Hence, managing a MarkLogic system required careful and continuous
% adjustment of these twins. For example, like any complex system,
% MarkLogic is controlled by numerous configuration files. It is an expert task
% to find the best settings to these configurations. Further, monitoring and controlling such a twin system
% requires considerable experience. 
% }

% \tosign[kevin]{For organizations without extensive systems experience in managing this kind of
% cloud architecture, it would be useful to have  automatic agents that can alert operators
% to any upcoming sub-optimal conditions.
% That's the motivation of this project -- predicting the system spikes and providing the system alarms in half an hour ahead so that the operational staffs  can get preparation.
% }

\section{Data Mining Technology}\label{sect:tech}

We adopted four widely used machine learning models as system spike learners, including logistic regression (LG), classification and regression trees (CART)/random forest(RF),
artificial neural network (ANN) and long-short term memory network (LSTM).
We chose  these four learners since  a recent survey of predicting service spike~\cite{syu2018survey} listed
LG, CART and ANN as the most common machine learning models to 
predict the web server response time.
They also suggest  several other  ML approaches applicable to time series modeling, including some more complicated ANNs developed and used in deep learning (e.g., LSTM).

% We also  applied two extra ML model enhancers in this project: SMOTE~\cite{chawla2002smote}
% and Hyperparameter Tuning"
% \bi
% \item
% As is shown in \fig{spike}, any day of data contains mostly non-spikes;
% i.e. our target class is relatively rare. Technically speaking, this means our training data
% is {\em highly imbalanced}. 
%  The synthetic minority over-sampling technique (or SMOTE), is a tool to mitigate the effects of data imbalance.
% \item
% Hyperparameter tuning strives to maximize a   model's predictive power  by setting various hyperparameter to the model.
% In recent years, more and more frameworks support   hyperparameter optimization
% including the Tensor-flow in Google cloud~\cite{google18}.
% \ei
% This section briefly introduced the above technology stack in the project {\it SPIKE}.  

\subsection{Logistic Regression}
Logistic regression analyzes the relationship between multiple independent variables and a categorical dependent variable, and estimates the probability of occurrence of an event by fitting data to a logistic curve \cite{Hyeoun}.
Two kinds  of logistic regression are binary or binomial logistic regression and multinomial logistic regression:
\bi
\item
\textit{Binary logistic regression} is used when the dependent variable is dichotomous and the independent variables are either continuous or categorical. 
\item
When the dependent variables are not dichotomous and is comprised of more than two categories, a \textit{multinomial logistic regression} can be employed.
\ei
Unlike linear regression, logistic regression can directly predict probabilities, i.e. the odds of a dependent variable happens.
The most essential assumption to the logistic regression is that the independent variables are linearly related to the log odds, that is the logit of the probability defined as 
$logit(p) = \log \frac{p}{1-p} (0<p<1)$.  

Logistic regression is used in various fields, including the system  maintenance.
For example, Hoffert {\it et al.}~\cite{hoffert2009using} trained the logistic regression models
to predict the response time  of a search and rescue (SAR) operations system and
therefore simplify the configuration of middleware and adaptive transport protocols.
%Hoeffert's work inspired us. We would see whether the logistic regression can predict the spikes in LexisNexis product environment.

\subsection{CART}
The decision tree, or specifically, the 
classification and regression trees \textit{CART}~\cite{rutkowski2014cart} is another common type of supervised learning algorithm. 
It works for both categorical and continuous input and output variables. CART splits the population or sample into two or more homogeneous sets (or sub-populations) based on most significant splitter/differentiator in input variables. The leaf nodes of the tree contain an dependent variable which is used to make a prediction.

There are two types of  trees:
\bi
\item
\textit{Classification Tree} which serves problems with categorical target variables;
\item
\textit{Regression Tree} which is applicable to problems with continuous target variables.
\ei
In our work,  when applying the decision tree, we  treated the target (service spikes indicator) as continuous target, instead of binary category.
An estimation of service spike values let operation team staffs have the sense how urgent the further actions should take. 

 In terms of generating comprehensible models, classification and regression trees are our preferred choice.
In our experience, if they can be kept under a few dozen nodes,
 decision/regression trees are fast to read and understand.
Using the trees, engineers can figure out the key factors that contribute the targets. That information is constructive in further system refactoring. For example, we can adjust more storage and bandwidth resource if the I/O is the key factor to system performance  to some microservice.

\subsection{Other Learners}

In order to assess the impact of our  
``comprehensability'' requirement  on learner performance, as described in this section, we also explored
some learning models that can produce somewhat opaque results.

{\it Random Forests}  construct multiple  trees at the training time. The prediction of such a forest
comes from the majority view of all its trees. 
Random forest was first introduced by Tin Kam Ho~\cite{ho1995random} and has been applied in many ML applications~\cite{alexander2014image,gromping2009variable,belgiu2016random}.
Random forests can produce a very large set of  trees which an be hard to read and understand.

Neural networks consist of input and output layers, as well as (in most cases) a hidden layer consisting of units that transform the input into something that the output layer can use \cite{dormehl}. 
The most basic type of neural net is something called a \textit{feedforward neural network}, in which information travels in only one direction from input to output. A more widely used 
type of network is the \textit{recurrent neural network} (RNN), in which data can flow in multiple directions. 

Neural network models represent their knowledge in a somewhat arcane distributed manner which is not human comprehensible.

\begin{figure}[!bt]
\includegraphics[width=8cm]{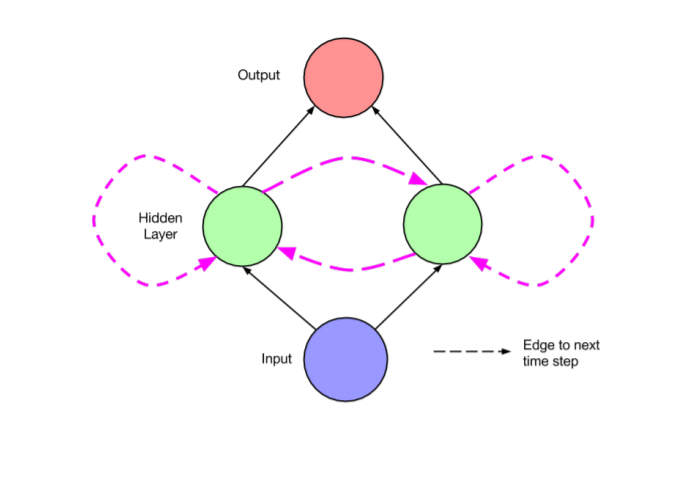}
\caption{A Simple Recurrent Neural Network \cite{lipton2015critical}}
\label{fig:rnn}
\end{figure}

Long-short term memory networks~\cite{hochreiter1997long}, or LSTM for short, is a special kind of RNN, capable of learning long-term dependencies.
\fig{rnn} shows a simple recurrent network. In RNN, the model is interpreted not as cyclic, but rather as a deep network with one layer per time step and shared weights across time steps. This algorithm is called back propagation through time~\cite{Werbos90}.
Recurrent Neural Networks suffer from problems
with short-term memory. During back propagation, recurrent neural networks suffer from the vanishing gradient problem
(gradients are values used to 
update a neural networks weights). The vanishing gradient problem is when the gradient shrinks as it back propagates through time. If a gradient value becomes extremely small, it does not contribute too much learning. LSTM was created as the solution to this short-term 
memory problem. The long short-term memory block is a complex unit with various components such as weighted inputs, activation functions, inputs from previous blocks and eventual outputs. 
The block is called a long short-term memory block because RNN is using a structure founded on short-term memory processes to create longer-term memory.

\begin{table*}[!t]
\caption {List of hyperparameters tuned.}
\label{tab:hyperparameters_opt}
\begin{tabular}{c|l|r|r|l}\hline
\rowcolor{black!10}
\textbf{Learner}&\textbf{Parameters}&\textbf{Default}&\textbf{Best}&\textbf{Descriptions}\\\hline
CART & min samples split & 2 & 0.61 & The minimum number of samples required to split an internal   node. \\ \cline{2-5}
     & max depth & None & 5 & The maximum depth of the tree. \\ \hline
RF & n\_estimators & 10 & 20 & The number of trees in the forest. \\\cline{2-5}
& min samples split & 2 & 0.1 & The minimum number of samples required to split an internal node. \\  \cline{2-5}
& max depth & None & 3 & The maximum depth of the tree. \\\hline
\end{tabular}
\end{table*}

\subsection{Data Pre-Processing with SMOTE}
As mentioned above, our training
data is very imbalanced. Specifically, 
from Nov 2018 to Jan 2019, the service spikes happened at only 3.4\% of the time.
Such imbalanced training data makes
it hard for classification models  to detect rare events~\cite{sun2009classification}.

There are several ways to apply
{\em resampling} to mitigate
for class imbalance~\cite{chawla2002smote,walden2014predicting,wallace2010semi,mani2003knn}:

\bi
\item
Over-sampling to make more of the minority class;
\item
Under-sampling to remove majority class items; 
\item
Some hybrid of the first two.  
\ei

Machine learning researchers~\cite{haixiang2017learning} advise that under-sampling can work better than over-sampling if there are hundreds of minority observations in the datasets. When there are only a few dozen minority instances, over-sampling approaches are superior to under-sampling. In the case of large size of training samples, the hybrid methods would be a better choice.

The Synthetic Minority Oversampling Technique(SMOTE)
\cite{chawla2002smote}
is a hybrid algorithm that performs both over- and under-sampling. SMOTE calculates the $k$ nearest neighbors for each minority class samples. Depending on the amount of oversampling required, one or more of the $k$-nearest neighbors are picked to create the synthetic samples. This amount is usually denoted by oversampling percentage (e.g., $50\%$ by default). The next step is to randomly creating a synthetic sample along the line connecting two minority samples.

\subsection{Parameter Tuning with Differential Evolution}
In machine learning, model hyperparameter are values in machine learning models that can require different constraints, weights or learning rates to generate different data patterns, e.g., the number of neighbours in $k$-Nearest Neighbours (KNN)~\cite{keller1985fuzzy}. 
Such hyperparameters are very important because they directly control the behaviors of the training algorithm and also impact the performance of the model being trained. Therefore, choosing appropriate hyperparameters plays a critical role in the performance of machine learning models. Hyperparameter tuning is the process of searching the most optimal hyperparameters for machine learning learners~\cite{biedenkapp2018hyperparameter,franceschi2017forward}.

Recent studies have shown that hyperparameter optimization can achieve better performance than using ``off-the-shelf'' configurations in several research areas in software engineering, e.g., software defect prediction~\cite{fu2016tuning, tosun2009reducing,osman2017hyperparameter,fu2017easy,krishna2017less} and software effort estimation~\cite{xia2018hyperparameter}. To the best of our knowledge, we are first to apply hyperparameter optimization in response time prediction.

Hyperparameter optimization can be implemented in many ways:
\bi
\item \textit{Grid search}~\cite{bergstra2011algorithms} loops through all combinations of all parameters. Although   Grid search  is a simple  to implement, it suffers if data have high dimensional space called the ``curse of dimensionality''. Previous work has shown that grid search might miss important optimization~\cite{fu2016differential} and is a time-wasting process since only a few of the tuning parameters really matters~\cite{bergstra2012random}.
% Grid search works by measuring performance using cross validation technique, e.g., K-Fold Cross Validation, which give assurance that our trained model gets most of the pattern from the data set. 
\item
\textit{Random search}~\cite{bergstra2012random} randomly samples the search space and evaluates sets from a specified probability distribution. Such 
random searches do not  use information from prior experiment to select the next set and also it is very difficult to predict the next set of experiments.
\item
\textit{Bayesian optimization}~\cite{pelikan1999boa} works by assuming the unknown function was sampled from a Gaussian Process and maintains a posterior distribution for this function as observation are made. However, it might be best-suited for optimization over continuous domains with small number of dimensions~\cite{frazier2018tutorial}.
\ei
This paper uses \textit{Differential evolution} (DE) for hyperparameter optimization.
DE has  proven useful in prior SE tuning
studies~\cite{fu2016tuning}. Also, our reading of the current literature is
that there are many advocates for differential evolution like Vesterstrom {\it et al.}~\cite{vesterstrom2004comparative} showed DE to be
competitive with particle swarm optimization and other genetic algorithms.

The premise of DE is that the best way to mutate the existing tunings is to extrapolate between current solutions. DE builds a population $P$ from a small number of randomly selected solutions of size $np$.
Then, each member of the population is compared against a  {\em mutant} built as follows.
Three solutions $(a, b, c) \in P$ are selected at random. For each tuning parameter $k$, at some probability $cr$, we replace the old tuning $x_k$ with $y_k$ as
% For
% booleans $y_k = \neg x_k$ and for numerics, 
\mbox{$y_k = a_k + f \times (b_k - c_k)$}
where $f$ is a parameter controlling differential weight.  
The main loop of DE runs over the population $P$ of size $np$, replacing old items with new candidates (if new candidate is better). This means that, as the loop progresses, the population is full of increasingly more valuable solutions (which, in turn,
helps   extrapolation).

For pragmatic reasons 
we did not tune all parameters of all learners.
LSTMs took 30 minutes to test each tuning. Given our DE settings,
that would have required 6000 hours of CPU; i.e. 25 weeks.
Table~\ref{tab:hyperparameters_opt} shows the parameters that we did tune. During that tuning process,
we asked our optimizers to maximize recall and precision.
As to the control parameters of DE,  using advice from Storn and Fu {\it et al.}~\cite{storn1997differential,fu2017easy}, we set $\{\mathit{np,f,cr}\}=\{20,0.75,0.3\}$. Also,
the number of generations $\mathit{gen}$ was set to 10 to test
the effects of  a very   CPU-light   optimizer.

For the set of parameters we did tune, see \tab{hyperparameters_opt}.

\section{Data}\label{sect:data}
% We spent around one month to understand the system architecture as well as goal of the project,  and then started to work on the {\it SPIKE} project.
% The exploration can be divided into 4 stages. The first two stages are the {\it data engineering} while the rest are the {\it data science}.
% \begin{enumerate}
% \item Monitor data collection and pre-processing;
% \item Data visualization;
% \item Real-time system performance prediction via common ML learners;
% \item Spike predictions for half an hour.
% \end{enumerate}

 For this analysis, we collected data from the  LexisNexis N-document
database searching microservice.
This  N-document database contains 20+ million documents.
% They are serviced by 
% 72 % note --> including the News and scitificnews
% AWS EC2 instances serving the customers searches and  queries.
%Each year the LexisNexis Group spends over  two millions dollars  in providing database query services,
%%supporting the legal consulting, risk management, etc.
%Most of such web services are built upon the MarkLogic services using Amazon Web Services (AWS) or Microsoft Azure.

\begin{figure*}[!t]
    \begin{tabular}{cc}
    \includegraphics[width=0.4\linewidth]{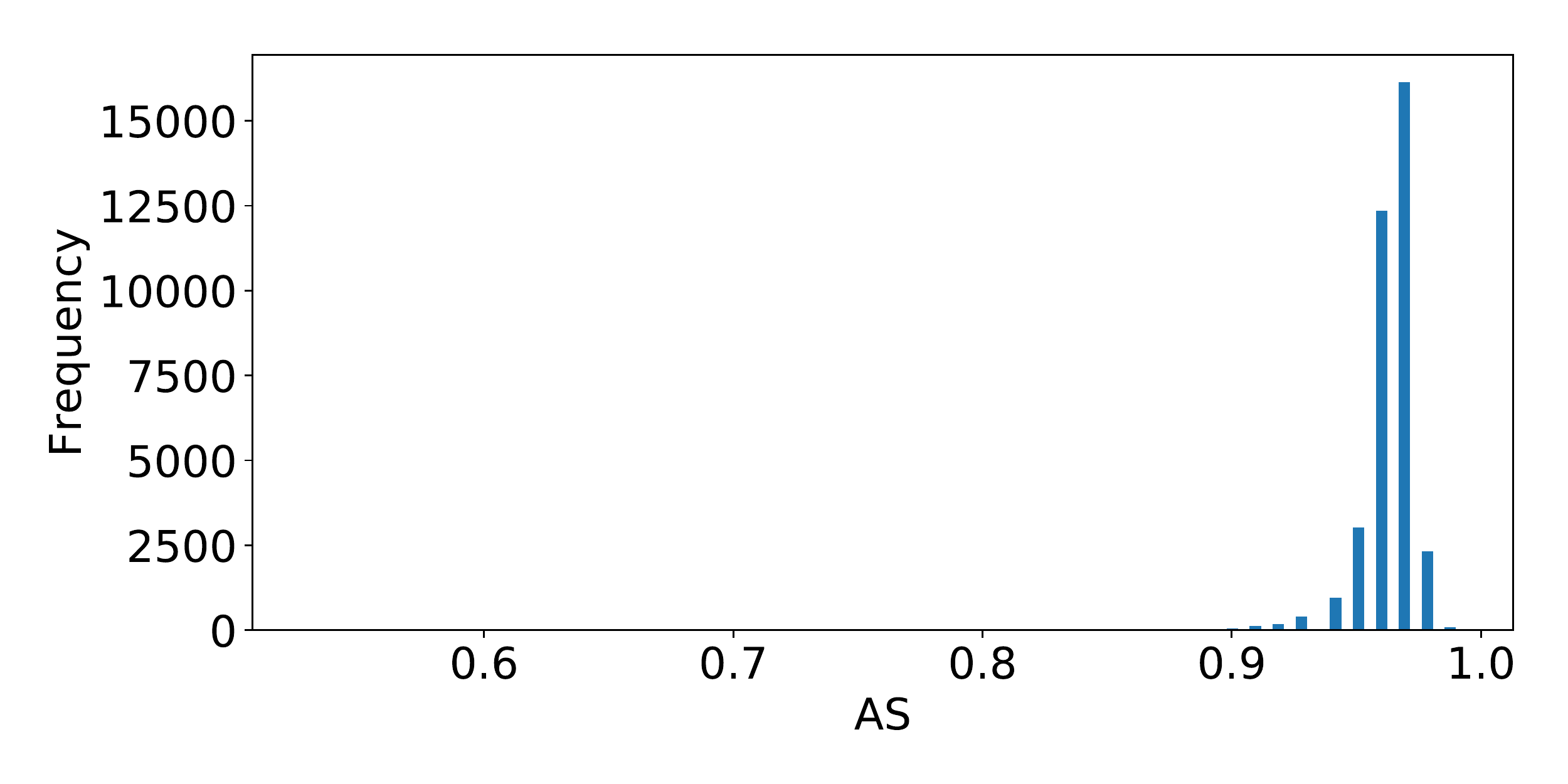} &
    \includegraphics[width=0.4\linewidth]{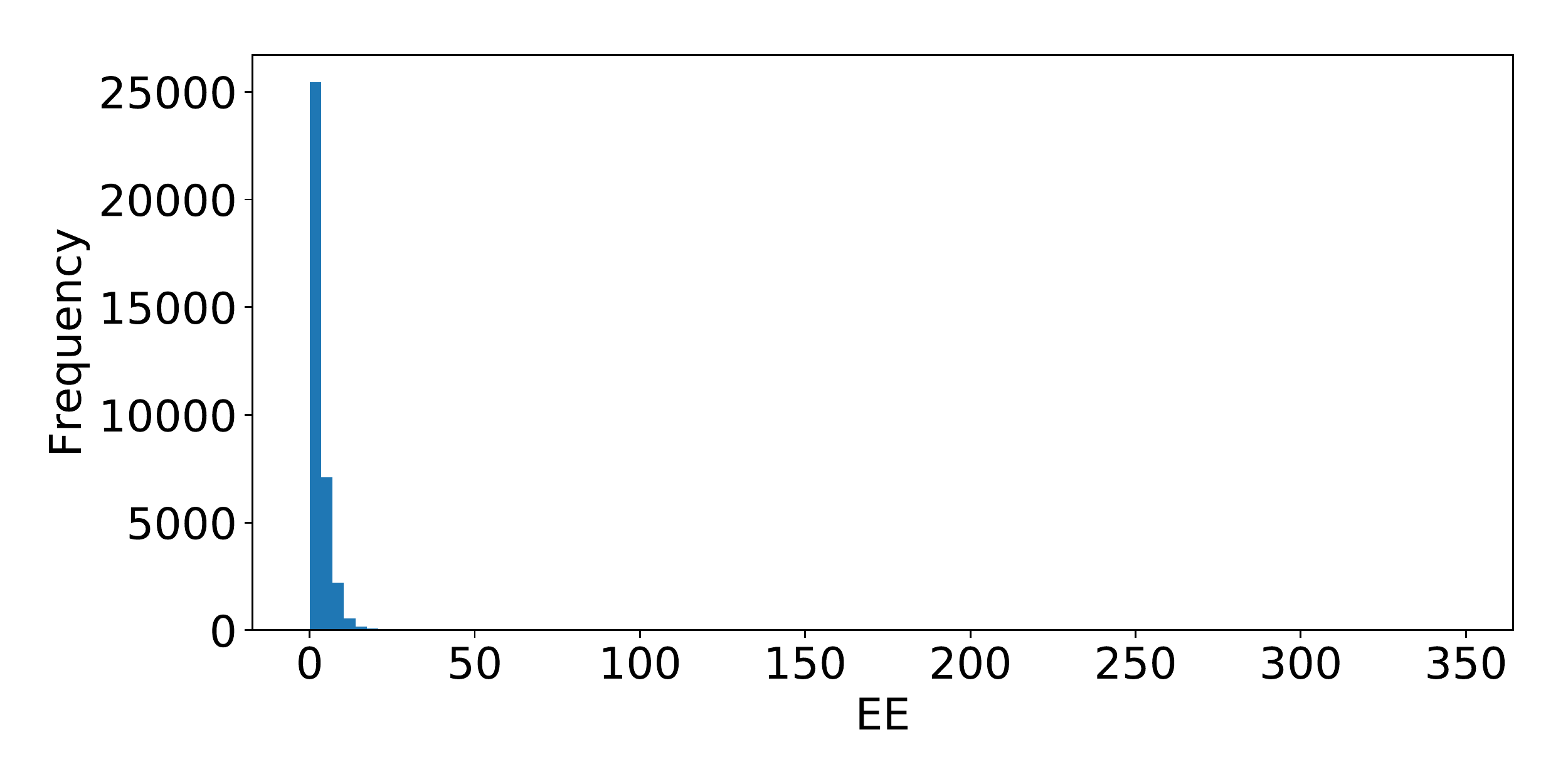}\\
    \includegraphics[width=0.4\linewidth]{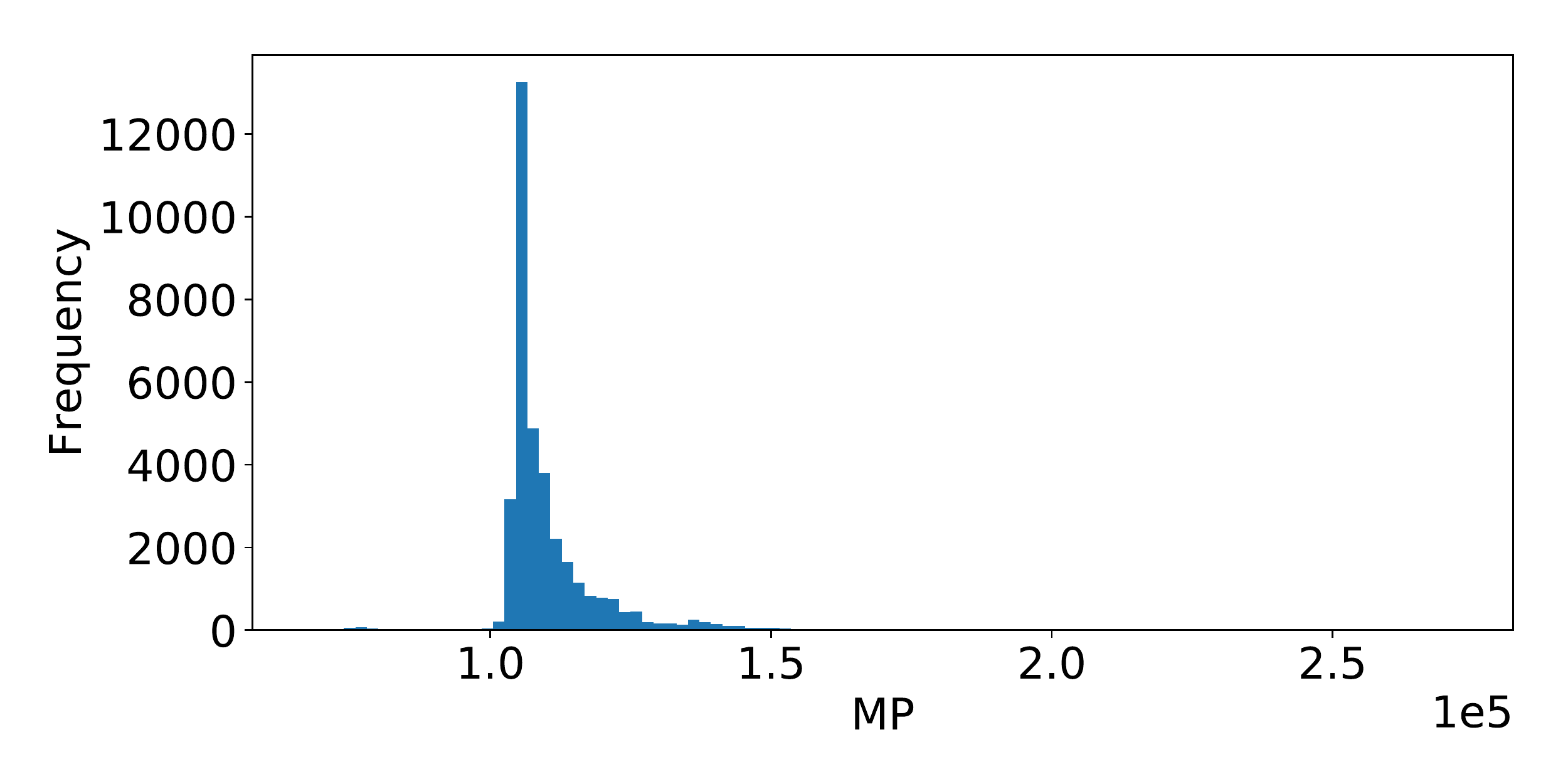}  &
    \includegraphics[width=0.4\linewidth]{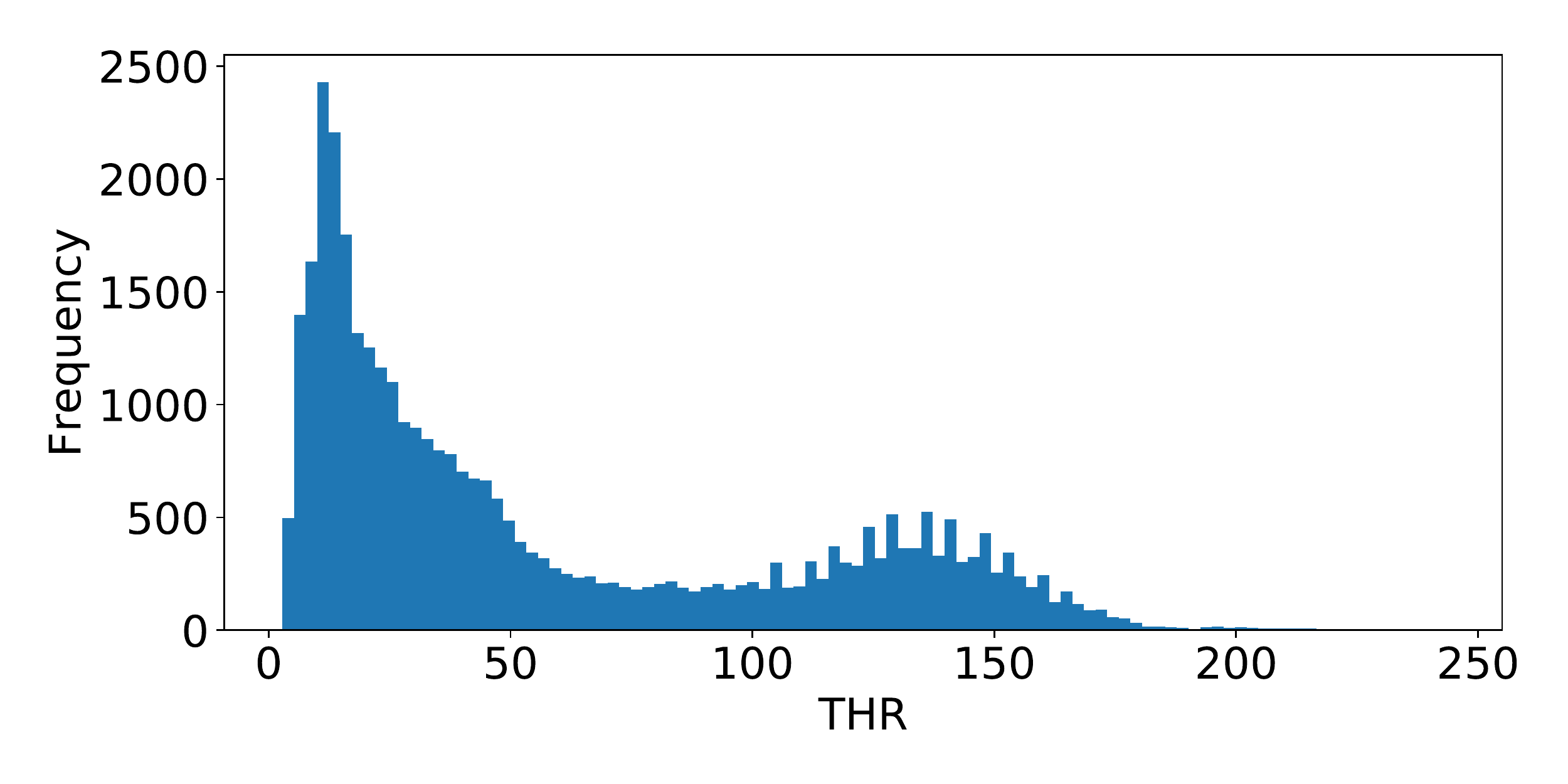}\\
    \end{tabular}
  \caption{Independent features histogram. $x$-axis shows sorted all monitoring values. $y$-axis is the corresponding frequency.}
  \label{fig:fea_hist}
\end{figure*}
\begin{figure}[!t]
  \includegraphics[width=.8\linewidth]{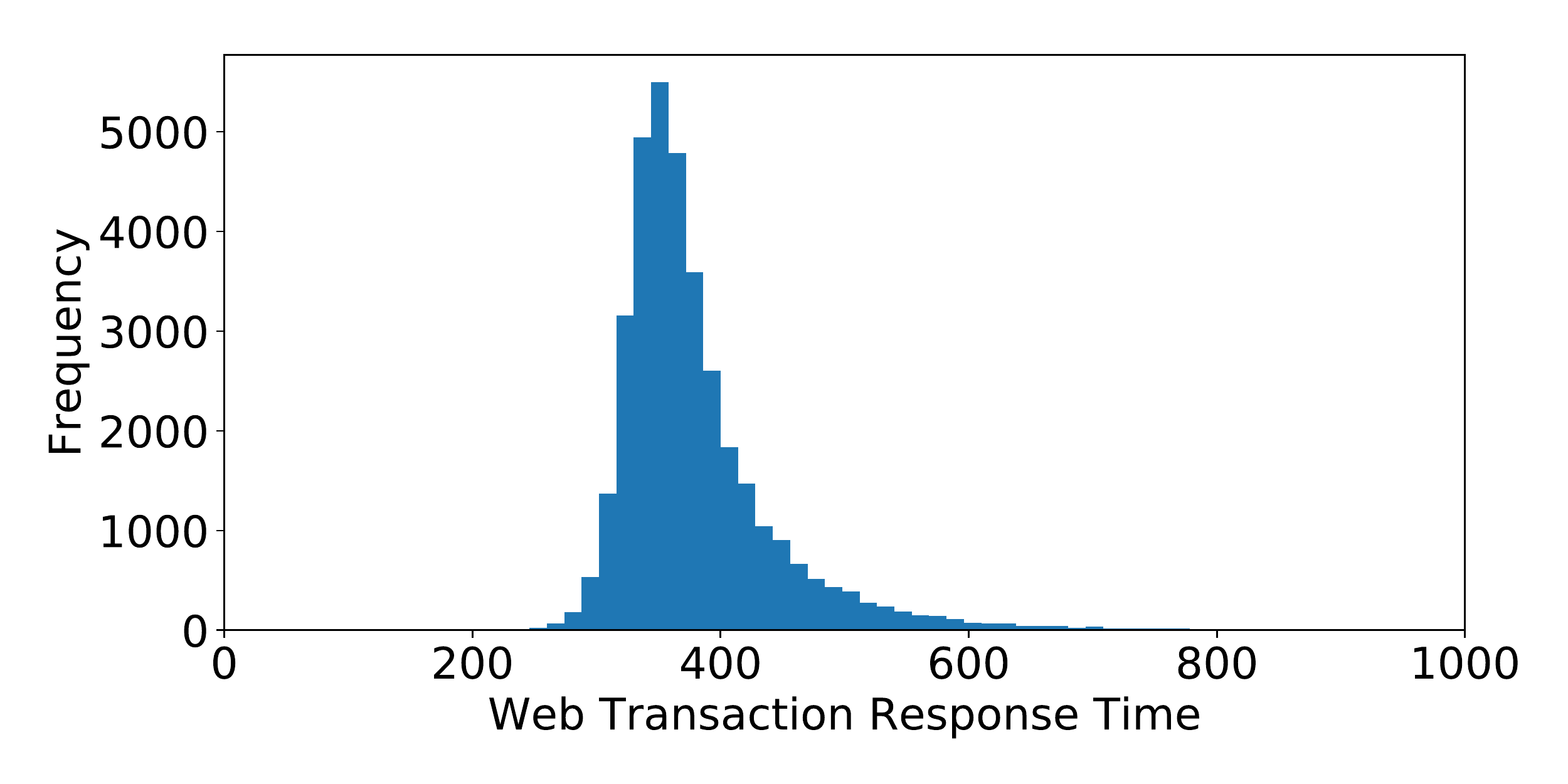}
  \caption{Histogram of \texttt{Web Service Response Time}. $x$-axis is the service response time in milliseconds. $y$-axis is the frequency in the training set. Note that this plot shows a  {\em dependent} variable that is {\em different} 
  to the THR$_t$ {\em independent}  variable discussed   in the text. }
  \label{fig:dep_hist}
\end{figure}

Recall from the introduction that the technology used in LexisNexis'
 cloud systems is changing rapidly. Accordingly, when we did data collection at one-minute intervals, and
 restricted
ourselves to  measurements we might reasonably expect to see in a wide range of future cloud environments.

% Specifically, we say that each node $x$ connects to multiple nodes $y_1,y_2,\ldots$   (and by ``connects'' we meant that these node $y_i$ reads or writes data from/to node $x$).
% That is, our training data has different columns for:
% \bi
% \item {\em Intra-node data} from each node $x$ in the cloud;
% \item {\em Inter-node data} that samples the topology of the network;
% i.e. that includes any information about  nodes $y$ that reads/writes   from/to  node $x$.
% \ei 

We collected data
on:
\bi
\item the current {\em web service response time} of the N-document
service;
\item that same response time at 
$t \in \{5, 10, 15, 30, 60, 90, 120, 150, $ $180, 300, 1440\}$ minutes into the past. %
\ei
We also collected the following data on the 
for each  13 microservices  directly upstream on the N-document
services
%created  5 columns of data showing   mean values seen in the  past 
%$t \in \{10, 30, 60, 300, 1440\}$ minutes:
\be
\item {\it Response time} for service $x$; 
\item {\it All Logged Errors Per Minute} (EE$_t$); 
\item {\it Total Physical Memory Used in MB} (MP$_t$);
\item {\it Web Transaction Throughput} (THR$_t$). 
\item The  {\it application performance index Apdex Score} (AS$_t$).   
The Apdex score is defined as 
the number of satisfied samples plus half of the tolerating samples plus none of the frustrated samples, divided by all the samples in unit minute. That is:
\begin{center}
{\em Apdex = (Satisfactory samples + 0.5$\times$Tolerating samples + 0$\times$Frustrated samples)/Total samples}
\end{center}
Here, {\em satisfactory, tolerating} and {\em frustrating} are defined in the standard way, as per~\cite{apdex}.
\ee
\begin{table}[!t]
\caption{Statistics of monitored data after the final steps in the pipeline.}
\label{tab:datastat}
\small
\begin{tabular}{l|rrrr} \hline
\rowcolor{black!10} \textbf{Percentile}& \textbf{EE} & \textbf{AS} & \textbf{MP} &       \textbf{THR} \\
\hline
%mean  &      2.94 &      0.96 &  109848 &     57.75 \\
%std   &      6.40 &      0.02 &    9581 &     50.91 \\
0\% (min)   &      0 &      0.53 &   67800 &      2.83 \\
25\%   &      0.12 &      0.96 &  105000 &     15.90 \\
50\%   &      1.72 &      0.97 &  107000 &     35.60 \\
75\%   &      3.88 &      0.97 &  111000 &    102 \\
100\% (max)   &    347 &      0.99 &  272000 &    243 \\ \hline
\end{tabular}
\end{table}
% For the {\em inter-node data}, for each node $y$ that connects to $x$, 
% we collected mean values for
% (EM$_t$, MP$_t$, THR$_t$, AS$_t$)\footnote{Some monitoring data may be missing due to system architecture reasons.}, 
% as well as their service response time,
% over the last $t=60$ minutes.
% For the case study of this paper, we were exploring a system of 12
% nodes. In total, our
% collected  data comprises 75 columns (variables):
% \bi
% \item
% From node $x$, there were 20 columns holding data from   5 time stamps
% of four variables  (EE$_t$, MP$_t$, TR$_t$, AS$_t$)  
% for  $t \in \{10, 30, 60, 300, 1440\}$ minutes;
% \item
% There was also 54 additional columns holding data for the 13 nodes  $y$ that read/write data
% from/to $x$. For each node $y$, we record the mean value for
%  (EE$_t$, MP$_t$, TR$_t$, AS$_t$) and service response time as seen in the last $t=60$ minutes. Please note that the response time in $y$ nodes are independent variables;
% \item The dependent column {\it web service response time} of $x$ node, i.e. the node serving the N-document search microservice.
% \ei
The columns of our collected data
had the distributions of \tab{datastat}.
 Figure \ref{fig:fea_hist} and \ref{fig:dep_hist} show the histograms of attributes(independent features) and class (dependent feature) respectively. 
From these \fig{fea_hist} and \ref{fig:dep_hist}, we have the following observations:
\bi
\item All our variables are highly non-evenly distributed, i.e. all of them have large standard deviation.
\item In statistics, a \textit{long tail} of some distributions of numbers is the portion of the distribution having a large number of occurrences far from the ``head'' or central part of the distribution \cite{longtail}. We see this pattern among all attributes.
\item From Figure \ref{fig:dep_hist} we can see that in majority time, the system response time varied from 0-470.  By selecting for \mbox{$x>470$}, we could focus this study on the most outstanding service spikes.
\ei

% \begin{figure}[!t]
% \centering
% \includegraphics[width=0.8\columnwidth]{resx_roc.pdf}
% \caption{ROC curve, corresponding to \fig{tor_pd}.
% }
% \label{fig:roc}
% \end{figure}

\section{Predicting with  SPIKE}
LexisNexis is serving customers from hundreds of areas. Their behavior pattern may be changing at all times, therefore lead to various situations of service spikes. For example, the way customers from Wall Street using the LexisNexis service is different to those from Silicon Valley -- at least, they are interested in distinct group of documents.
As a result, model to predict "financial news" spikes is not applicable to predict "tech scandal" spikes.
Therefore it is important to train  {\it SPIKE} on recent local
data that is specific to a particular web
service.

That said, deciding what data mining method to apply is a 
time-consuming and CPU expensive process. Therefore, our work 
was divided
into two stages:
\be
\item {\em Model selection and tuning} where we pruned options
to find one promising model. For this stage, we used 
one month of data divided into an 80\% train  and   20\% test
phase (where the test data was selected from the  last week of
the month).
\item
{\em Testing our most promising method}. For this stage, we used
a different month of data to test the method selected during
stage one.
We stepped through this data in ``windows'' of ten minutes.
At each step $i$, we trained the model (found by {\em model selection and tuning}) using the next
24 hours of data (i.e. windows $i$ to $i+144$).
This model was then tested using data from the next half hour of data
(i.e. windows $i+j, j\in\{145,146,147\}$).
This means that the stage2 results (reported below) come from 4317 different train/test pairs.
\ee
For details on these two stages, see below.

\newcolumntype{H}{>{\setbox0=\hbox\bgroup}c<{\egroup}@{}}
\begin{table*}[!t]
\caption{Learning on first 80\% of data, testing on the more recent 20\%. Results sorted by recall (and {\em higher} values are {\em better}).
ANN=Artificial Neural network, LSTM=Long-short Term Memory. 
In this table {\em higher} values for recall, precision.
}
\label{tab:res1}
\begin{tabular}{rccccccrr}
   \multicolumn{4}{r}{$TP$}  & $FP$ & $FN$   & $TN$    & $r = \frac{TP}{TP+FN}$ & $p = \frac{TP}{TP+FP}$ \vspace{1mm}\\
\rowcolor{black!70} \textbf{\textcolor{white}{Learner}}& \textbf{\textcolor{white}{SMOTE?}} & \textbf{\textcolor{white}{TUNE?}} & \textbf{\textcolor{white}{TP}}    & \textbf{\textcolor{white}{FP}} & \textbf{\textcolor{white}{FN}}   & \textbf{\textcolor{white}{TN}} & \textbf{\textcolor{white}{Recall(\%)}} & \textbf{\textcolor{white}{Precision(\%)}}  
% \textbf{\textcolor{white}{F-measure}}
\\
\rowcolor{black!20}ANN(10$\times$4)	&		&		&	504	&	1756	&	27	&	799	&	95	&	22			\\
\rowcolor{black!20} Regression Tree (CART)	&	\checkmark	&	\checkmark	&	493	&	868	&	38	&	1687	&	93	&	36			\\
\rowcolor{black!20}Random Forest	&	\checkmark	&		&	489	&	764	&	42	&	1791	&	92	&	39			\\
Random Forest	&		&		&	428	&	439	&	103	&	2116	&	81	&	49		\\
 Random Forest	&	\checkmark	&	\checkmark	&	417	&	437	&	114	&	2118	&	79	&	49	\\
Decision Tree (CART)	&	\checkmark	&		&	294	&	432	&	237	&	2123	&	55	&	40		\\
Decision Tree (CART)	&		&		&	232	&	362	&	299	&	2193	&	44	&	39			\\
LSTM	&	\checkmark	&		&	53	&	78	&	478	&	2477	&	10	&	40	\\
Logistic Regression	&	\checkmark	&		&	42	&	36	&	489	&	2519	&	8	&	54		\\
ANN(~5$\times$4)	&	\checkmark	&		&	40	&	44	&	491	&	2511	&	8	&	48			\\
ANN(~5$\times$4)	&		&		&	35	&	12	&	496	&	2543	&	7	&	74			\\
Logistic Regression	&		&		&	31	&	9	&	500	&	2546	&	6	&	77			\\
LSTM	&		&		&	30	&	23	&	501	&	2532	&	6	&	57		\\
ANN(10$\times$4)	&	\checkmark	&		&	31	&	33	&	500	&	2522	&	6	&	48	\\	 
\hline 
\end{tabular}
~\\

\end{table*}

\begin{figure*}[!b]
\centering
\includegraphics[width=\linewidth]{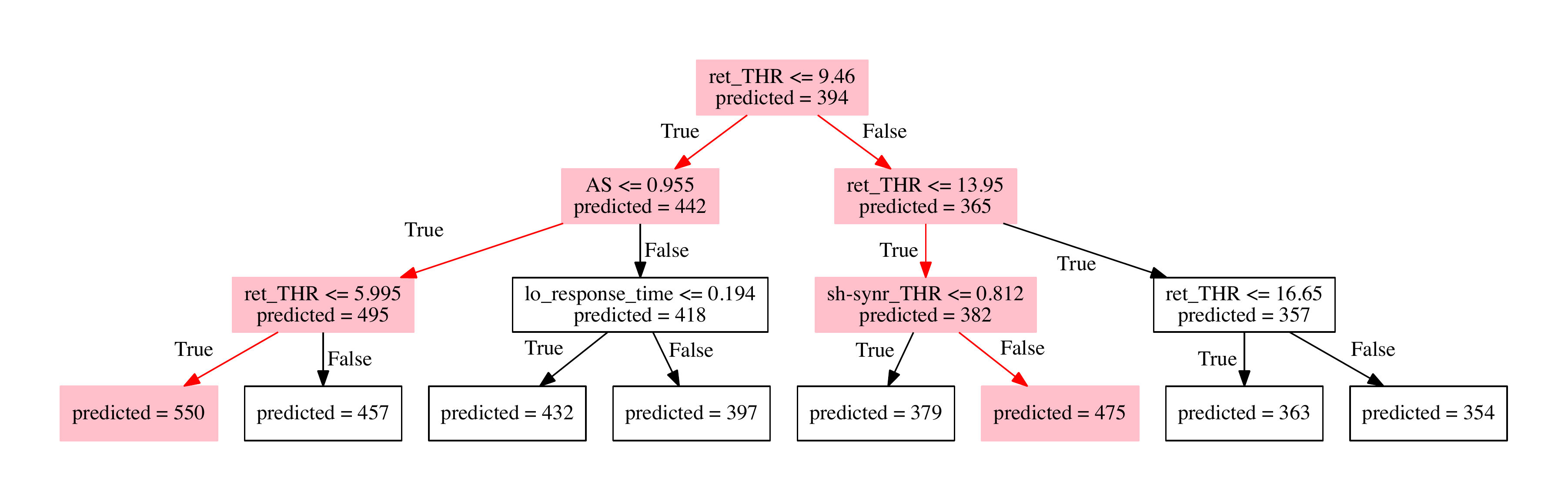}
\caption{Model generated from CART
\texttt{ret}, \texttt{lo} and \texttt{sh-synr} are 
node names.  
Branches leading to spikes (with service response time $>$470ms/query) are \color{red!70}{highlighted in red}.
}\label{fig:dt}
\end{figure*}

\subsection{Stage One: Model Selection and Tuning}

In this stage,
we sorted our one month data by time then trained on the first 80\% and tested on the last 20\%
\footnote{We did not use a randomized strategy to produce train/test sets since it makes more methodological sense to predict the future from the past.}.
The models were trained to get the real-time {\it web service response time} in this stage. 

Not all treatments were applied to all data sets. For example, as mentioned
above, the neural nets were too slow to tune.  As for the other learners, when optimizing hyperparameters, at the request of our business users,
we optimized for maximizing recall. Recall is defined
for a two-class classifier so if this stage, to guide the
hyperparameter optimization, we defined ``spike'' as per
Figure~\ref{fig:dep_hist}; i.e. greater than 470ms.

Important point:   SMOTE or hyperparameter optimization never used information from the test data.
If we  applied SMOTE or  hyperparameter optimization, these algorithms were used {\em only} on the training data;
i.e. our results are not over-fitted to  test data.

In all,
we explored the  14 treatments of \tab{res1}. 
As shown by the rows highlighted in gray, three methods achieved very high recalls of over 90\% (Random Forests, CART, ANN).
Initially, off-the-shelf CART performed poorly.
However, when augmented with   SMOTE and tuning, CART achieved very high recalls of over 90\%
(the associated precisions are not good-- which is a problem solved by the sensitivity analysis of the next section).

Applying the {\em comprehensibility } criteria, our summary of \tab{res1} is that 
SMOTE+ Regression Trees (CART) + hyperparameter optimization performs best.  
 While ANN offered marginally better recalls, 
 it is hard to read those models. 
CART, on the other hand, produced the  simple regression tree
of Figure~\ref{fig:dt}.

 Since this tree is easy to comprehend, it is easy to extract important business knowledge about LexisNexis cloud machines. For example:
 \bi
 \item
 Among all 13 nodes studied, only three were found to be important by this tree: \texttt{ret}, \texttt{lo} and \texttt{sh-synr}.  Prior to this study, the importance of these nodes to healthy operations at LexisNexis has not been realized. 
 \item
 To avoid
 spikes, engineers are advised to    take action that avoids
 the the red branches of \fig{dt}.
 \ei
 
\begin{figure*}[!t]
\begin{minipage}{3.2in}
\includegraphics[width=3in]{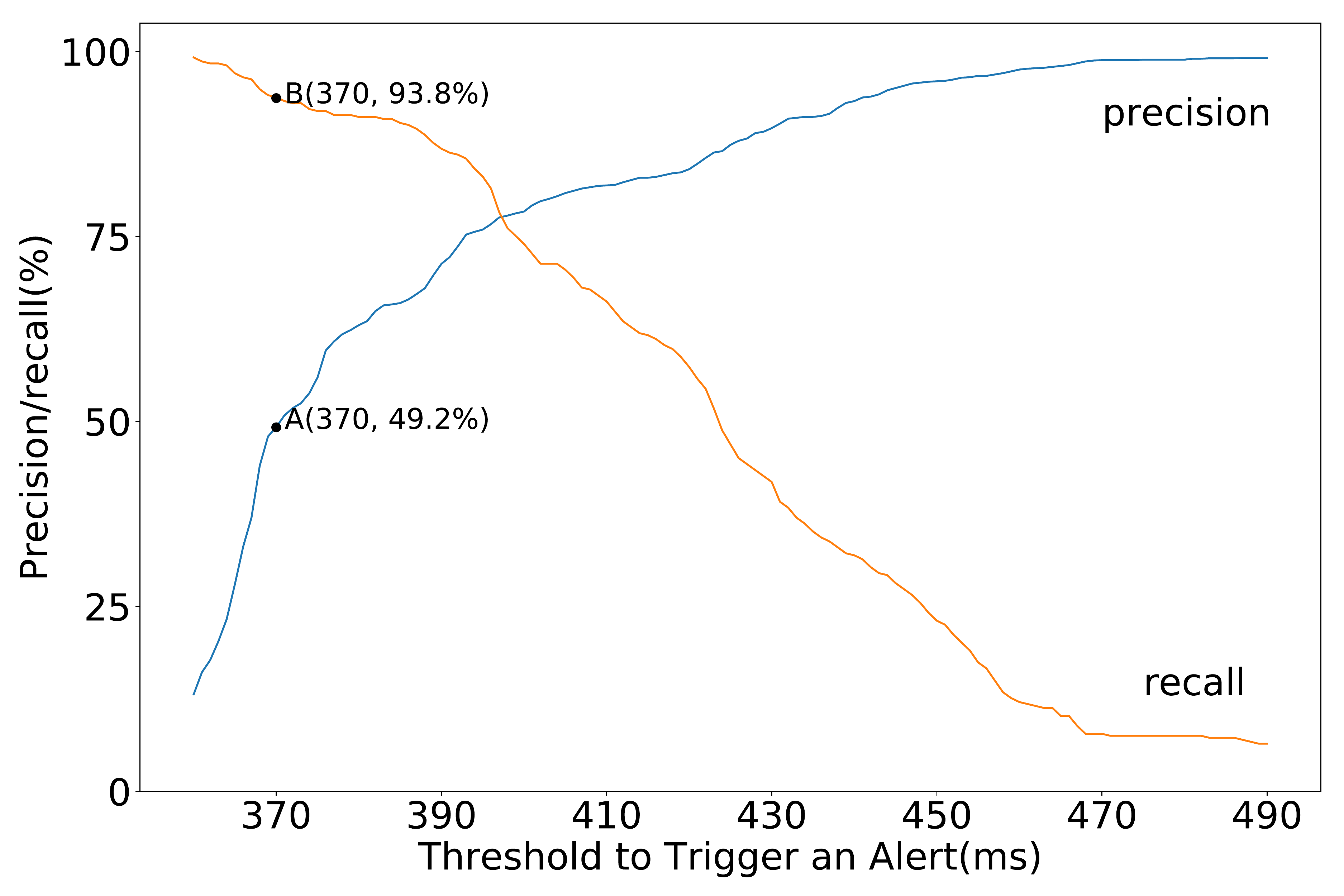}
\end{minipage}~\begin{minipage}{3in}
To understand the precision and recalls, consider the threshold $370ms$ that triggers an alert: 
\bi
\item
Among all actual future spikes(response time $>470ms$), 49.2\% of them have predicted value $>370ms$ (therefore trigger the alarm), which shows as point $A$; 
\item
Among all alarms triggered (i.e. predicted future response time $>370ms$), 93.8\% of them did exceed the threshold (having response time $>470ms$), indicated as point $B$.
\ei
\end{minipage}
\caption{Precision and recalls under different sensitivity to predict services spikes half an hour ahead. $x$-axis is the threshold value used  to trigger an alarm. 
% \fig{roc} shows the corresponding ROC curve.
}
\label{fig:tor_pd}
\end{figure*}

\subsection{Stage Two: Testing our Most Promising Method}\label{sect:pred_ahead}

Stage one found  that the  tree learner (CART) with specific hyperparameters (as shown in \tab{hyperparameters_opt}) was comparatively
better than several other methods. 

This second stage tests if that model is useful on real-world data.
To predict the {\it web service response time} at the moment (t+.5)hr,
{\it SPIKE} trained the data from [$t-L$hrs, t),
where $L$ is the training window size. We explored   
\mbox{$L \in \{0.25, 0.5,1,5, 24, 48\}$} and found
that best results come from training from
 the last $L=24$hrs of prior data.

We found that the precisions and recalls achieved by CART+ SMOTE+ optimization were sensitive to our threshold
for recognizing a spike. Hence, we show results
where we declared a  ``spike'' being defined as more than
from 370ms to 490ms. 

\fig{tor_pd} shows results seen
while adjusting the threshold for predicting a spike.
As shown in this figure, at a threshold of 404ms,
can achieve  precisions and recalls of  75\% or higher.  

It is worth mentioning that, according to the experiment, there always exist some false positives or 
false negatives.
The impact of false positive and false negative signals from the on the business is generally limited, but it can grow to devalue to the tool.  
High levels of false negative signals would result in no action taken by support engineers prior to service incidents.  
This would essentially be ``business as usual'', and likely not a negative impact other than decreasing confidence in the tool. 
On the other hand, high levels of false positive signals would cause support engineers to engage in monitoring activities prematurely, often diverting them from other tasks.  
While this seems like a high cost, discussions with LexisNexis
site reliability engineering team indicates that this cost is acceptable. 
The time it takes to open an application performance monitoring tool to view current status is a negligible cost in comparison to the benefit of potentially avoiding a customer impacting service degradation.

\section{Conclusions and Lessons Learned}

Using regression trees (e.g CART)
and synthetic over-sampling of rare events (e.g. via SMOTE)
and hyperparameter optimization (e.g. using DE),
 optimizers it is possible to build effective and comprehensive
predictors for service spikes.  {\it SPIKE} can predict with reasonable recall and precisions if a spike will occur in the next 30   minutes.  Further, {\it SPIKE} can report its reasoning via
 a very small and easily comprehensible    tree, from which we can learn important (and previously unknown) aspects about this domain.

The factors that lead to service spikes are highly context specific. Much time was spent in this work trying solutions from other sites~\cite{syu2018survey}, which proved less-than-satisfactory for this problem (neural nets, logistic regression).
 There exist tools for exploring a large number of options within data miners.
If we had our time again, we would {\em first} commission a hyperparameter optimization (a tool to explore
all those options) before {\em secondly}
use those optimizers to faster explore different data mining options.

 But even with hyperparameter optimizers, building   predictors is a complex tasks
 (certainly, much more than running one    query, then glancing at a simple data dashboard).
 There is considerable creativity required in how to design the inputs to a learning problem and how 
 to find tune the resulting models. For example, in this work, we made poor progress until
 we somewhat serendipitously decided to: 
 \bi
 \item Add inter-node information to the training set (see Section~4);
 \item Conduct  a sensitivity  analysis (see   previous section).
 \ei
 More generally,
a modern cloud environment can generate petabytes of operational logs, every day. 
For example,
LexisNexis  constantly
monitors  the state of its cloud services,
collecting data from many  microservices at one-minute intervals.
A data science team exploring the problem of service
spikes needs considerable business knowledge to ``slice and dice''   the data. 
In all,  the  results of this paper took three months
to generate:
\bi
\item 1 month of a LexisNexis data engineer generating our training data by writing complex joins across large datasets.
\item
1 month of
inductive engineering, applying  different data mining methods to the data.
As mentioned above,
this proved to be a  tedious task that required  developed and discarding a dozen very bad predictors before 
finding
one that achieved useful results
\item 1 month of a senior LexisNexis engineer serving as a liaison between our team and the rest of LexisNexis.
The importance of the liaison   cannot be overstated. 
That person (a)~maintained senior management's awareness and enthusiasm for this project;
(b)~organized access to   numerous subject matter experts. 
\ei
When staffing similar efforts in the future, we recommend a similar ``three-sided''
team comprising 
inductive engineers, data engineers, and   business knowledge experts.

\section{Future Work}

We believe {\it SPIKE} is a general method for managing rare,
but critical, CPU issues
in complex cloud environments:
\be
\item
For each node, 
{\it SPIKE} trains models using
(a)~{\em intra-node} details about the recent history of
that node as well as (b)~some {\em inter-node} knowledge about connected nodes.
\item
For  rare events, it is important to
use class rebalancing tools (like SMOTE). 
\item
Also, since the factors that lead to
service spikes are highly context specific, it is useful to employ
hyperparameter optimization (like DE).
\ee
In theory, these  three  principles  should apply to
other services at LexisNexis and other organizations. 
In future work we aim to test that conjecture using more data.

Furthermore, most tools SRE's deal with either have an ``Alarm'' state or an ``Ok'' state.  This type of tool is more a ``probability'' tool of a potential alarm state (defined as the threshold like \fig{tor_pd}). It is an interesting future work to call out a ``confidence'' that something is going to go into an ``Alarm'' state.

Also, after {\em prediction} comes {\em diagnosis} and 
{\em repair}. If we build
trees like Figure~\ref{fig:dt} from
more data (covering more months and more LexisNexis services) then 
we would be able to uncover critical thresholds for critical nodes
that most effect LexisNexis services. Using that knowledge, plus
more subject matter expertise, we should then be able to propose
spike reduction policies.

\section*{Acknowledgements}
This work was partially funded by (a)~a gift from LexisNexis
managed by Phillpe Poignant; and (b)~an NSF CCF grant \#1703487.

\balance
\bibliographystyle{ACM-Reference-Format}

\begin{thebibliography}{00}

%%% the Web address.

\ifx \showCODEN    \undefined \def \showCODEN     #1{\unskip}     \fi
\ifx \showDOI      \undefined \def \showDOI       #1{#1}\fi
\ifx \showISBNx    \undefined \def \showISBNx     #1{\unskip}     \fi
\ifx \showISBNxiii \undefined \def \showISBNxiii  #1{\unskip}     \fi
\ifx \showISSN     \undefined \def \showISSN      #1{\unskip}     \fi
\ifx \showLCCN     \undefined \def \showLCCN      #1{\unskip}     \fi
\ifx \shownote     \undefined \def \shownote      #1{#1}          \fi
\ifx \showarticletitle \undefined \def \showarticletitle #1{#1}   \fi
\ifx \showURL      \undefined \def \showURL       {\relax}        \fi
% invisible to TeX
\providecommand\bibfield[2]{#2}
\providecommand\bibinfo[2]{#2}
\providecommand\natexlab[1]{#1}
\providecommand\showeprint[2][]{arXiv:#2}

\bibitem[\protect\citeauthoryear{??}{lnn}{1994}]%
        {lnnyt1}
 \bibinfo{year}{1994}\natexlab{}.
\newblock \showarticletitle{Company News; A Name Change is Planned for Mead
  Data Central}.
\newblock \bibinfo{journal}{{\em The New York Times.\/}}
  (\bibinfo{year}{1994}).
\newblock
\showURL{%
\url{https://www.nytimes.com/1994/12/02/business/company-news-a-name-change-is-planned-for-mead-data-central.html?src=pm}}


\bibitem[\protect\citeauthoryear{??}{lnl}{2019}]%
        {lnlegal}
 \bibinfo{year}{2019}\natexlab{}.
\newblock \showarticletitle{LEGAL}.
\newblock \bibinfo{journal}{{\em www.relx.com\/}} (\bibinfo{year}{2019}).
\newblock
\showURL{%
\url{https://www.relx.com/our-business/market-segments/legal}}


\bibitem[\protect\citeauthoryear{??}{apd}{2019}]%
        {apdex}
 \bibinfo{year}{2019 (accessed Apr 1, 2019)}\natexlab{}.
\newblock \bibinfo{booktitle}{{\em Apdex Alliance}}.
\newblock
\newblock
\shownote{\url{http://www.apdex.org/}.}


\bibitem[\protect\citeauthoryear{Alexander, Zikic, Zhang, Zhang, and
  Criminisi}{Alexander et~al\mbox{.}}{2014}]%
        {alexander2014image}
\bibfield{author}{\bibinfo{person}{Daniel~C Alexander}, \bibinfo{person}{Darko
  Zikic}, \bibinfo{person}{Jiaying Zhang}, \bibinfo{person}{Hui Zhang}, {and}
  \bibinfo{person}{Antonio Criminisi}.} \bibinfo{year}{2014}\natexlab{}.
\newblock \showarticletitle{Image quality transfer via random forest
  regression: applications in diffusion MRI}. In \bibinfo{booktitle}{{\em
  International Conference on Medical Image Computing and Computer-Assisted
  Intervention}}. Springer, \bibinfo{pages}{225--232}.
\newblock


\bibitem[\protect\citeauthoryear{Belgiu and Dr{\u{a}}gu{\c{t}}}{Belgiu and
  Dr{\u{a}}gu{\c{t}}}{2016}]%
        {belgiu2016random}
\bibfield{author}{\bibinfo{person}{Mariana Belgiu} {and}
  \bibinfo{person}{Lucian Dr{\u{a}}gu{\c{t}}}.}
  \bibinfo{year}{2016}\natexlab{}.
\newblock \showarticletitle{Random forest in remote sensing: A review of
  applications and future directions}.
\newblock \bibinfo{journal}{{\em ISPRS Journal of Photogrammetry and Remote
  Sensing\/}}  \bibinfo{volume}{114} (\bibinfo{year}{2016}),
  \bibinfo{pages}{24--31}.
\newblock


\bibitem[\protect\citeauthoryear{Bergstra and Bengio}{Bergstra and
  Bengio}{2012}]%
        {bergstra2012random}
\bibfield{author}{\bibinfo{person}{James Bergstra} {and}
  \bibinfo{person}{Yoshua Bengio}.} \bibinfo{year}{2012}\natexlab{}.
\newblock \showarticletitle{Random search for hyper-parameter optimization}.
\newblock \bibinfo{journal}{{\em Journal of Machine Learning Research\/}}
  \bibinfo{volume}{13}, \bibinfo{number}{Feb} (\bibinfo{year}{2012}),
  \bibinfo{pages}{281--305}.
\newblock


\bibitem[\protect\citeauthoryear{Bergstra, Bardenet, Bengio, and
  K{\'e}gl}{Bergstra et~al\mbox{.}}{2011}]%
        {bergstra2011algorithms}
\bibfield{author}{\bibinfo{person}{James~S Bergstra}, \bibinfo{person}{R{\'e}mi
  Bardenet}, \bibinfo{person}{Yoshua Bengio}, {and} \bibinfo{person}{Bal{\'a}zs
  K{\'e}gl}.} \bibinfo{year}{2011}\natexlab{}.
\newblock \showarticletitle{Algorithms for hyper-parameter optimization}. In
  \bibinfo{booktitle}{{\em Advances in neural information processing systems}}.
  \bibinfo{pages}{2546--2554}.
\newblock


\bibitem[\protect\citeauthoryear{Biedenkapp, Eggensperger, Elsken, Falkner,
  Feurer, Gargiani, Hutter, Klein, Lindauer, Loshchilov,
  et~al\mbox{.}}{Biedenkapp et~al\mbox{.}}{2018}]%
        {biedenkapp2018hyperparameter}
\bibfield{author}{\bibinfo{person}{Andre Biedenkapp},
  \bibinfo{person}{Katharina Eggensperger}, \bibinfo{person}{Thomas Elsken},
  \bibinfo{person}{Stefan Falkner}, \bibinfo{person}{Matthias Feurer},
  \bibinfo{person}{Matilde Gargiani}, \bibinfo{person}{Frank Hutter},
  \bibinfo{person}{Aaron Klein}, \bibinfo{person}{Marius Lindauer},
  \bibinfo{person}{Ilya Loshchilov}, {et~al\mbox{.}}}
  \bibinfo{year}{2018}\natexlab{}.
\newblock \showarticletitle{Hyperparameter Optimization}.
\newblock \bibinfo{journal}{{\em Artificial Intelligence\/}}
  \bibinfo{volume}{1} (\bibinfo{year}{2018}), \bibinfo{pages}{35}.
\newblock


\bibitem[\protect\citeauthoryear{Bingham and Spradlin}{Bingham and
  Spradlin}{2011}]%
        {longtail}
\bibfield{author}{\bibinfo{person}{Alpheus Bingham} {and}
  \bibinfo{person}{Dwayne Spradlin}.} \bibinfo{year}{2011}\natexlab{}.
\newblock \showarticletitle{The Long Tail of Expertise}.
\newblock \bibinfo{journal}{{\em Pearson Education\/}}  \bibinfo{volume}{ISBN
  9780132823135} (\bibinfo{year}{2011}).
\newblock
\showURL{%
\url{https://books.google.com/books?id=9qLfsonmwhAC&pg=PT5#v=onepage&q&f=false}}


\bibitem[\protect\citeauthoryear{Chawla, Bowyer, Hall, and Kegelmeyer}{Chawla
  et~al\mbox{.}}{2002}]%
        {chawla2002smote}
\bibfield{author}{\bibinfo{person}{Nitesh~V Chawla}, \bibinfo{person}{Kevin~W
  Bowyer}, \bibinfo{person}{Lawrence~O Hall}, {and} \bibinfo{person}{W~Philip
  Kegelmeyer}.} \bibinfo{year}{2002}\natexlab{}.
\newblock \showarticletitle{SMOTE: synthetic minority over-sampling technique}.
\newblock \bibinfo{journal}{{\em Journal of artificial intelligence
  research\/}}  \bibinfo{volume}{16} (\bibinfo{year}{2002}),
  \bibinfo{pages}{321--357}.
\newblock


\bibitem[\protect\citeauthoryear{Dormehl}{Dormehl}{2019}]%
        {dormehl}
\bibfield{author}{\bibinfo{person}{Luke Dormehl}.}
  \bibinfo{year}{2019}\natexlab{}.
\newblock \showarticletitle{What is an artificial neural network? Here\'s
  everything you need to know}.
\newblock \bibinfo{journal}{{\em EMERGING TECH\/}} (\bibinfo{year}{2019}).
\newblock
\showURL{%
\url{https://www.digitaltrends.com/cool-tech/what-is-an-artificial-neural-network/}}


\bibitem[\protect\citeauthoryear{Franceschi, Donini, Frasconi, and
  Pontil}{Franceschi et~al\mbox{.}}{2017}]%
        {franceschi2017forward}
\bibfield{author}{\bibinfo{person}{Luca Franceschi}, \bibinfo{person}{Michele
  Donini}, \bibinfo{person}{Paolo Frasconi}, {and}
  \bibinfo{person}{Massimiliano Pontil}.} \bibinfo{year}{2017}\natexlab{}.
\newblock \showarticletitle{Forward and reverse gradient-based hyperparameter
  optimization}.
\newblock \bibinfo{journal}{{\em arXiv preprint arXiv:1703.01785\/}}
  (\bibinfo{year}{2017}).
\newblock


\bibitem[\protect\citeauthoryear{Frazier}{Frazier}{2018}]%
        {frazier2018tutorial}
\bibfield{author}{\bibinfo{person}{Peter~I Frazier}.}
  \bibinfo{year}{2018}\natexlab{}.
\newblock \showarticletitle{A tutorial on bayesian optimization}.
\newblock \bibinfo{journal}{{\em arXiv preprint arXiv:1807.02811\/}}
  (\bibinfo{year}{2018}).
\newblock


\bibitem[\protect\citeauthoryear{Fu and Menzies}{Fu and Menzies}{2017}]%
        {fu2017easy}
\bibfield{author}{\bibinfo{person}{Wei Fu} {and} \bibinfo{person}{Tim
  Menzies}.} \bibinfo{year}{2017}\natexlab{}.
\newblock \showarticletitle{Easy over hard: A case study on deep learning}. In
  \bibinfo{booktitle}{{\em Proceedings of the 2017 11th Joint Meeting on
  Foundations of Software Engineering}}. ACM, \bibinfo{pages}{49--60}.
\newblock


\bibitem[\protect\citeauthoryear{Fu, Menzies, and Shen}{Fu
  et~al\mbox{.}}{2016a}]%
        {fu2016tuning}
\bibfield{author}{\bibinfo{person}{Wei Fu}, \bibinfo{person}{Tim Menzies},
  {and} \bibinfo{person}{Xipeng Shen}.} \bibinfo{year}{2016}\natexlab{a}.
\newblock \showarticletitle{Tuning for software analytics: Is it really
  necessary?}
\newblock \bibinfo{journal}{{\em Information and Software Technology\/}}
  \bibinfo{volume}{76} (\bibinfo{year}{2016}), \bibinfo{pages}{135--146}.
\newblock


\bibitem[\protect\citeauthoryear{Fu, Nair, and Menzies}{Fu
  et~al\mbox{.}}{2016b}]%
        {fu2016differential}
\bibfield{author}{\bibinfo{person}{Wei Fu}, \bibinfo{person}{Vivek Nair}, {and}
  \bibinfo{person}{Tim Menzies}.} \bibinfo{year}{2016}\natexlab{b}.
\newblock \showarticletitle{Why is Differential Evolution Better than Grid
  Search for Tuning Defect Predictors?}
\newblock \bibinfo{journal}{{\em arXiv preprint arXiv:1609.02613\/}}
  (\bibinfo{year}{2016}).
\newblock


\bibitem[\protect\citeauthoryear{Gr{\"o}mping}{Gr{\"o}mping}{2009}]%
        {gromping2009variable}
\bibfield{author}{\bibinfo{person}{Ulrike Gr{\"o}mping}.}
  \bibinfo{year}{2009}\natexlab{}.
\newblock \showarticletitle{Variable importance assessment in regression:
  linear regression versus random forest}.
\newblock \bibinfo{journal}{{\em The American Statistician\/}}
  \bibinfo{volume}{63}, \bibinfo{number}{4} (\bibinfo{year}{2009}),
  \bibinfo{pages}{308--319}.
\newblock


\bibitem[\protect\citeauthoryear{Haixiang, Yijing, Shang, Mingyun, Yuanyue, and
  Bing}{Haixiang et~al\mbox{.}}{2017}]%
        {haixiang2017learning}
\bibfield{author}{\bibinfo{person}{Guo Haixiang}, \bibinfo{person}{Li Yijing},
  \bibinfo{person}{Jennifer Shang}, \bibinfo{person}{Gu Mingyun},
  \bibinfo{person}{Huang Yuanyue}, {and} \bibinfo{person}{Gong Bing}.}
  \bibinfo{year}{2017}\natexlab{}.
\newblock \showarticletitle{Learning from class-imbalanced data: Review of
  methods and applications}.
\newblock \bibinfo{journal}{{\em Expert Systems with Applications\/}}
  \bibinfo{volume}{73} (\bibinfo{year}{2017}), \bibinfo{pages}{220--239}.
\newblock


\bibitem[\protect\citeauthoryear{Ho}{Ho}{1995}]%
        {ho1995random}
\bibfield{author}{\bibinfo{person}{Tin~Kam Ho}.}
  \bibinfo{year}{1995}\natexlab{}.
\newblock \showarticletitle{Random decision forests}. In
  \bibinfo{booktitle}{{\em Proceedings of 3rd international conference on
  document analysis and recognition}}, Vol.~\bibinfo{volume}{1}. IEEE,
  \bibinfo{pages}{278--282}.
\newblock


\bibitem[\protect\citeauthoryear{Hochreiter and Schmidhuber}{Hochreiter and
  Schmidhuber}{1997}]%
        {hochreiter1997long}
\bibfield{author}{\bibinfo{person}{Sepp Hochreiter} {and}
  \bibinfo{person}{J{\"u}rgen Schmidhuber}.} \bibinfo{year}{1997}\natexlab{}.
\newblock \showarticletitle{Long short-term memory}.
\newblock \bibinfo{journal}{{\em Neural computation\/}} \bibinfo{volume}{9},
  \bibinfo{number}{8} (\bibinfo{year}{1997}), \bibinfo{pages}{1735--1780}.
\newblock


\bibitem[\protect\citeauthoryear{Hoffert, Mack, and Schmidt}{Hoffert
  et~al\mbox{.}}{2009}]%
        {hoffert2009using}
\bibfield{author}{\bibinfo{person}{Joe Hoffert}, \bibinfo{person}{Daniel Mack},
  {and} \bibinfo{person}{Douglas Schmidt}.} \bibinfo{year}{2009}\natexlab{}.
\newblock \showarticletitle{Using machine learning to maintain pub/sub system
  qos in dynamic environments}. In \bibinfo{booktitle}{{\em Proceedings of the
  8th international workshop on adaptive and reflective middleware}}. ACM,
  \bibinfo{pages}{4}.
\newblock


\bibitem[\protect\citeauthoryear{Keller, Gray, and Givens}{Keller
  et~al\mbox{.}}{1985}]%
        {keller1985fuzzy}
\bibfield{author}{\bibinfo{person}{James~M Keller}, \bibinfo{person}{Michael~R
  Gray}, {and} \bibinfo{person}{James~A Givens}.}
  \bibinfo{year}{1985}\natexlab{}.
\newblock \showarticletitle{A fuzzy k-nearest neighbor algorithm}.
\newblock \bibinfo{journal}{{\em IEEE transactions on systems, man, and
  cybernetics\/}} \bibinfo{number}{4} (\bibinfo{year}{1985}),
  \bibinfo{pages}{580--585}.
\newblock


\bibitem[\protect\citeauthoryear{Krishna, Menzies, and Layman}{Krishna
  et~al\mbox{.}}{2017}]%
        {krishna2017less}
\bibfield{author}{\bibinfo{person}{Rahul Krishna}, \bibinfo{person}{Tim
  Menzies}, {and} \bibinfo{person}{Lucas Layman}.}
  \bibinfo{year}{2017}\natexlab{}.
\newblock \showarticletitle{Less is more: Minimizing code reorganization using
  XTREE}.
\newblock \bibinfo{journal}{{\em Information and Software Technology\/}}
  \bibinfo{volume}{88} (\bibinfo{year}{2017}), \bibinfo{pages}{53--66}.
\newblock


\bibitem[\protect\citeauthoryear{Lipton, Berkowitz, and Elkan}{Lipton
  et~al\mbox{.}}{2015}]%
        {lipton2015critical}
\bibfield{author}{\bibinfo{person}{Zachary~C. Lipton}, \bibinfo{person}{John
  Berkowitz}, {and} \bibinfo{person}{Charles Elkan}.}
  \bibinfo{year}{2015}\natexlab{}.
\newblock \bibinfo{title}{A Critical Review of Recurrent Neural Networks for
  Sequence Learning}.
\newblock   (\bibinfo{year}{2015}).
\newblock
\showeprint[arxiv]{cs.LG/1506.00019}


\bibitem[\protect\citeauthoryear{Mani and Zhang}{Mani and Zhang}{2003}]%
        {mani2003knn}
\bibfield{author}{\bibinfo{person}{Inderjeet Mani} {and} \bibinfo{person}{I
  Zhang}.} \bibinfo{year}{2003}\natexlab{}.
\newblock \showarticletitle{kNN approach to unbalanced data distributions: a
  case study involving information extraction}. In \bibinfo{booktitle}{{\em
  Proceedings of workshop on learning from imbalanced datasets}},
  Vol.~\bibinfo{volume}{126}.
\newblock


\bibitem[\protect\citeauthoryear{Miller}{Miller}{2012}]%
        {lnnyt2}
\bibfield{author}{\bibinfo{person}{Stephen Miller}.}
  \bibinfo{year}{2012}\natexlab{}.
\newblock \showarticletitle{For Future Reference, a Pioneer in Online Reading}.
\newblock \bibinfo{journal}{{\em The Wall Street Journal\/}}
  (\bibinfo{year}{2012}).
\newblock
\showURL{%
\url{https://www.wsj.com/articles/SB10001424052970203721704577157211501855648?KEYWORDS=lexisnexis}}


\bibitem[\protect\citeauthoryear{Osman, Ghafari, and Nierstrasz}{Osman
  et~al\mbox{.}}{2017}]%
        {osman2017hyperparameter}
\bibfield{author}{\bibinfo{person}{Haidar Osman}, \bibinfo{person}{Mohammad
  Ghafari}, {and} \bibinfo{person}{Oscar Nierstrasz}.}
  \bibinfo{year}{2017}\natexlab{}.
\newblock \showarticletitle{Hyperparameter optimization to improve bug
  prediction accuracy}. In \bibinfo{booktitle}{{\em Machine Learning Techniques
  for Software Quality Evaluation (MaLTeSQuE), IEEE Workshop on}}. IEEE,
  \bibinfo{pages}{33--38}.
\newblock


\bibitem[\protect\citeauthoryear{Park}{Park}{2013}]%
        {Hyeoun}
\bibfield{author}{\bibinfo{person}{Hyeoun-Ae Park}.}
  \bibinfo{year}{2013}\natexlab{}.
\newblock \showarticletitle{An Introduction to Logistic Regression: From Basic
  Concepts to Interpretation with Particular Attention to Nursing Domain}.
\newblock \bibinfo{journal}{{\em Korean Society of Nursing Science\/}}
  \bibinfo{volume}{43}, \bibinfo{number}{2} (\bibinfo{year}{2013}),
  \bibinfo{pages}{1--5}.
\newblock
\showURL{%
\url{http://synapse.koreamed.org/DOIx.php?id=10.4040%2Fjkan.2013.43.2.154}}


\bibitem[\protect\citeauthoryear{Pelikan, Goldberg, and Cant{\'u}-Paz}{Pelikan
  et~al\mbox{.}}{1999}]%
        {pelikan1999boa}
\bibfield{author}{\bibinfo{person}{Martin Pelikan}, \bibinfo{person}{David~E
  Goldberg}, {and} \bibinfo{person}{Erick Cant{\'u}-Paz}.}
  \bibinfo{year}{1999}\natexlab{}.
\newblock \showarticletitle{BOA: The Bayesian optimization algorithm}. In
  \bibinfo{booktitle}{{\em Proceedings of the 1st Annual Conference on Genetic
  and Evolutionary Computation-Volume 1}}. Morgan Kaufmann Publishers Inc.,
  \bibinfo{pages}{525--532}.
\newblock


\bibitem[\protect\citeauthoryear{Rutkowski, Jaworski, Pietruczuk, and
  Duda}{Rutkowski et~al\mbox{.}}{2014}]%
        {rutkowski2014cart}
\bibfield{author}{\bibinfo{person}{Leszek Rutkowski}, \bibinfo{person}{Maciej
  Jaworski}, \bibinfo{person}{Lena Pietruczuk}, {and} \bibinfo{person}{Piotr
  Duda}.} \bibinfo{year}{2014}\natexlab{}.
\newblock \showarticletitle{The CART decision tree for mining data streams}.
\newblock \bibinfo{journal}{{\em Information Sciences\/}}
  \bibinfo{volume}{266} (\bibinfo{year}{2014}), \bibinfo{pages}{1--15}.
\newblock


\bibitem[\protect\citeauthoryear{Storn and Price}{Storn and Price}{1997}]%
        {storn1997differential}
\bibfield{author}{\bibinfo{person}{Rainer Storn} {and} \bibinfo{person}{Kenneth
  Price}.} \bibinfo{year}{1997}\natexlab{}.
\newblock \showarticletitle{Differential evolution--a simple and efficient
  heuristic for global optimization over continuous spaces}.
\newblock \bibinfo{journal}{{\em Journal of global optimization\/}}
  \bibinfo{volume}{11}, \bibinfo{number}{4} (\bibinfo{year}{1997}),
  \bibinfo{pages}{341--359}.
\newblock


\bibitem[\protect\citeauthoryear{Sun, Wong, and Kamel}{Sun
  et~al\mbox{.}}{2009}]%
        {sun2009classification}
\bibfield{author}{\bibinfo{person}{Yanmin Sun}, \bibinfo{person}{Andrew~KC
  Wong}, {and} \bibinfo{person}{Mohamed~S Kamel}.}
  \bibinfo{year}{2009}\natexlab{}.
\newblock \showarticletitle{Classification of imbalanced data: A review}.
\newblock \bibinfo{journal}{{\em International Journal of Pattern Recognition
  and Artificial Intelligence\/}} \bibinfo{volume}{23}, \bibinfo{number}{04}
  (\bibinfo{year}{2009}), \bibinfo{pages}{687--719}.
\newblock


\bibitem[\protect\citeauthoryear{Syu, Wang, and Fanjiang}{Syu
  et~al\mbox{.}}{2018}]%
        {syu2018survey}
\bibfield{author}{\bibinfo{person}{Yang Syu}, \bibinfo{person}{Chien-Min Wang},
  {and} \bibinfo{person}{Yong-Yi Fanjiang}.} \bibinfo{year}{2018}\natexlab{}.
\newblock \showarticletitle{A Survey of Time-Aware Dynamic QoS Forecasting
  Research, Its Future Challenges and Research Directions}. In
  \bibinfo{booktitle}{{\em International Conference on Services Computing}}.
  Springer, \bibinfo{pages}{36--50}.
\newblock


\bibitem[\protect\citeauthoryear{Tosun and Bener}{Tosun and Bener}{2009}]%
        {tosun2009reducing}
\bibfield{author}{\bibinfo{person}{Ayse Tosun} {and} \bibinfo{person}{Ayse
  Bener}.} \bibinfo{year}{2009}\natexlab{}.
\newblock \showarticletitle{Reducing false alarms in software defect prediction
  by decision threshold optimization}. In \bibinfo{booktitle}{{\em Proceedings
  of the 2009 3rd International Symposium on Empirical Software Engineering and
  Measurement}}. IEEE Computer Society, \bibinfo{pages}{477--480}.
\newblock


\bibitem[\protect\citeauthoryear{Vance}{Vance}{2010}]%
        {lnnyt}
\bibfield{author}{\bibinfo{person}{Ashlee Vance}.}
  \bibinfo{year}{2010}\natexlab{}.
\newblock \showarticletitle{Legal Sites Plan Revamps as Rivals Undercut Price}.
\newblock \bibinfo{journal}{{\em The New York Times.\/}}
  (\bibinfo{year}{2010}).
\newblock
\showURL{%
\url{https://www.nytimes.com/2010/01/25/technology/25westlaw.html?_r=1&ref=reedelsevier}}


\bibitem[\protect\citeauthoryear{Vesterstr{\o}m and Thomsen}{Vesterstr{\o}m and
  Thomsen}{2004}]%
        {vesterstrom2004comparative}
\bibfield{author}{\bibinfo{person}{Jakob Vesterstr{\o}m} {and}
  \bibinfo{person}{Rene Thomsen}.} \bibinfo{year}{2004}\natexlab{}.
\newblock \showarticletitle{A comparative study of differential evolution,
  particle swarm optimization, and evolutionary algorithms on numerical
  benchmark problems}. In \bibinfo{booktitle}{{\em Evolutionary Computation,
  2004. CEC2004. Congress on}}, Vol.~\bibinfo{volume}{2}. IEEE,
  \bibinfo{pages}{1980--1987}.
\newblock


\bibitem[\protect\citeauthoryear{Walden, Stuckman, and Scandariato}{Walden
  et~al\mbox{.}}{2014}]%
        {walden2014predicting}
\bibfield{author}{\bibinfo{person}{James Walden}, \bibinfo{person}{Jeff
  Stuckman}, {and} \bibinfo{person}{Riccardo Scandariato}.}
  \bibinfo{year}{2014}\natexlab{}.
\newblock \showarticletitle{Predicting vulnerable components: Software metrics
  vs text mining}. In \bibinfo{booktitle}{{\em Software Reliability Engineering
  (ISSRE), 2014 IEEE 25th International Symposium on}}. IEEE,
  \bibinfo{pages}{23--33}.
\newblock


\bibitem[\protect\citeauthoryear{Wallace, Trikalinos, Lau, Brodley, and
  Schmid}{Wallace et~al\mbox{.}}{2010}]%
        {wallace2010semi}
\bibfield{author}{\bibinfo{person}{Byron~C Wallace}, \bibinfo{person}{Thomas~A
  Trikalinos}, \bibinfo{person}{Joseph Lau}, \bibinfo{person}{Carla Brodley},
  {and} \bibinfo{person}{Christopher~H Schmid}.}
  \bibinfo{year}{2010}\natexlab{}.
\newblock \showarticletitle{Semi-automated screening of biomedical citations
  for systematic reviews}.
\newblock \bibinfo{journal}{{\em BMC bioinformatics\/}} \bibinfo{volume}{11},
  \bibinfo{number}{1} (\bibinfo{year}{2010}), \bibinfo{pages}{55}.
\newblock


\bibitem[\protect\citeauthoryear{{Werbos}}{{Werbos}}{1990}]%
        {Werbos90}
\bibfield{author}{\bibinfo{person}{P.~J. {Werbos}}.}
  \bibinfo{year}{1990}\natexlab{}.
\newblock \showarticletitle{Backpropagation through time: what it does and how
  to do it}.
\newblock \bibinfo{journal}{{\it Proc. IEEE}} \bibinfo{volume}{78},
  \bibinfo{number}{10} (\bibinfo{date}{Oct} \bibinfo{year}{1990}),
  \bibinfo{pages}{1550--1560}.
\newblock
\showISSN{0018-9219}
\showDOI{%
\url{https://doi.org/10.1109/5.58337}}


\bibitem[\protect\citeauthoryear{Xia, Krishna, Chen, Mathew, Shen, and
  Menzies}{Xia et~al\mbox{.}}{2018}]%
        {xia2018hyperparameter}
\bibfield{author}{\bibinfo{person}{Tianpei Xia}, \bibinfo{person}{Rahul
  Krishna}, \bibinfo{person}{Jianfeng Chen}, \bibinfo{person}{George Mathew},
  \bibinfo{person}{Xipeng Shen}, {and} \bibinfo{person}{Tim Menzies}.}
  \bibinfo{year}{2018}\natexlab{}.
\newblock \showarticletitle{Hyperparameter Optimization for Effort Estimation}.
\newblock \bibinfo{journal}{{\em arXiv preprint arXiv:1805.00336\/}}
  (\bibinfo{year}{2018}).
\newblock


\end{thebibliography}

\end{document}